\documentclass[12pt]{article}
\usepackage{amsmath}
\usepackage{amssymb}
\usepackage{epsfig}
\usepackage{euscript}
\usepackage{fancybox}
\usepackage{color}

\def\mathswitchr#1{\relax\ifmmode{\mathrm{#1}}\else$\mathrm{#1}$\fi}

\newcommand {\pslash}{\hbox{$\not\hbox{\kern-2.3pt $p$}$}}

\usepackage{cite}

\usepackage{epic}

\def\alf1{ {\alpha\over\pi} }

\begin{document}
\begin{titlepage}
\begin{flushright}
{\bf BU-HEPP-10-04 }\\
{\bf Jul., 2010}\\
\end{flushright}
 
\begin{center}
{\Large An Estimate of $\Lambda$ in Resummed Quantum Gravity
in the Context of Asymptotic Safety$^{\dagger}$
}
\end{center}

\vspace{2mm}
\begin{center}
{\bf   B.F.L. Ward}\\
\vspace{2mm}
{\em Department of Physics,\\
 Baylor University, Waco, Texas, 76798-7316, USA}\\
\end{center}

\vspace{5mm}
\begin{center}
{\bf   Abstract}
\end{center}
We show that, by using recently developed exact resummation techniques
based on the extension of the methods of Yennie, Frautschi and Suura
to Feynman's formulation of Einstein's theory, we get quantum field theoretic
descriptions for the UV fixed-point behaviors of the dimensionless
gravitational and cosmological constants postulated by Weinberg. Connecting
our work to
the attendant phenomenological asymptotic safety analysis of Planck scale
cosmology by Bonanno and Reuter, we estimate the value of the cosmological constant $\Lambda$. We find the encouraging estimate $\rho_\Lambda\equiv \frac{\Lambda}{8\pi G_N} \simeq (2.4\times 10^{-3}eV)^4$. 
While this numerical value is close to
recent experimental observations, we caution the reader that the estimate involves a number of model parameters that still possess significant levels of uncertainty, such as the value of the transition time between the Planck scale cosmology era and the Friedmann-Robertson-Walker radiation dominated era, where our current understanding allows for at least two orders of magnitude
in its uncertainty and this would change our estimate of $\rho_\Lambda$ by at least four orders of magnitude. We discuss such theoretical uncertainties as well.
We show why GUT and EW scale vacuum energies from spontaneous 
symmetry breaking
are suppressed in our approach to the estimation of $\rho_\Lambda$.
As a bonus, we show how our estimate constrains susy GUTS.
\vspace{10mm}
\vspace{10mm}
\renewcommand{\baselinestretch}{0.1}
\footnoterule
\noindent
{\footnotesize
\begin{itemize}
\item[${\dagger}$]
Work partly supported
by NATO Grant PST.CLG.980342.
\end{itemize}
}

\end{titlepage}

\def\Kmax{K_{\rm max}}\def\ieps{{i\epsilon}}\def\rQCD{{\rm QCD}}
\renewcommand{\theequation}{\arabic{equation}}
\font\fortssbx=cmssbx10 scaled \magstep2
\renewcommand\thepage{}
\parskip.1truein\parindent=20pt\pagenumbering{arabic}\par
\section{\bf Introduction}\label{intro}
\par
In Ref.~\cite{wein1}, Weinberg suggested that the general theory of relativity may have a non-trivial UV fixed point, with a finite dimensional critical surface
in the UV limit, so that it would be asymptotically safe with an S-matrix
that depends on only a finite number of observable parameters. 
In Refs.~\cite{reutera,laut,reuterb,reuter3,litim,perc}, strong evidence has been calculated
using Wilsonian~\cite{kgw} field-space exact renormalization group methods to support
Weinberg's asymptotic safety hypothesis for the Einstein-Hilbert theory.
As we review briefly below, in a parallel but independent development~\cite{bw1,bw2,bw2a,bw2b,bw2c,bw2d,bw2e,bw2f,bw2g,bw2h}, we have shown~\cite{bw2i} that the extension of the amplitude-based, exact resummation theory of Ref.~\cite{yfs, yfs-jw} to the Einstein-Hilbert theory leads to UV-fixed-point behavior for the dimensionless
gravitational and cosmological constants, but with the added bonus that the resummed theory is actually UV finite when expanded in the resummed propagators and vertices to any finite order in the respective improved loop expansion.
We have called the resummed theory resummed quantum gravity. 
More recently, more evidence for Weinberg's asymptotic safety behavior has been calculated using causal dynamical triangulated lattice methods in Ref.~\cite{ambj}\footnote{We also note that the model in Ref.~\cite{horva} realizes many aspects
of the effective field theory implied by the anomalous dimension of 2 at the
UV-fixed point but it does so at the expense of violating Lorentz invariance.}.
At this point, there is no known inconsistency between our analysis
and those of the Refs.~\cite{reutera,laut,reuterb,reuter3,litim,perc,ambj}.\par
We need to stress that the results in Refs.~\cite{reutera,laut,reuterb,reuter3,litim,perc}, while impressive, involve cut-offs which remain in the results
to varying degrees even for products such as that for the UV limits of the 
dimensionless gravitational and cosmological constants. 
In addition, the results
in Refs.~\cite{reutera,laut,reuterb,reuter3,litim,perc} retain some 
mild dependence on gauge parameters, again even for the product of 
the UV limits 
of the dimensionless
gravitational and cosmological constants. 
Accordingly, henceforward, we refer to the approach in 
Refs.~\cite{reutera,laut,reuterb,reuter3,litim,perc} as the 
'phenomenological' asymptotic safety approach.
What can be said is that 
dependencies are mild enough that the existence of the non-Gaussian UV 
fixed point found in these references is probably a physical result. 
But, until a 
rigorously cut-off independent and gauge invariant calculation corroborates 
these results, we cannot consider them final. Our approach offers such a 
calculation, as our results are both gauge invariant and cut-off independent. 
The results from Refs.~\cite{ambj}, involving, as they most certainly do, 
lattice constant-type artifact 
issues, are also only an indication of what the true continuum limit 
might realize -- they too need to be corroborated by a rigorous calculation 
without the issues of finite size and other possible lattice artifacts to 
be considered final. Again, our approach offers an answer to these issues.
The stage is therefore prepared for us to try to make contact with experiment, 
as such contact is the ultimate purpose of theoretical physics.\par
Toward this end, we note that, in Refs.~\cite{reuter1,reuter2}, 
it has been argued that the attendant phenomenological
asymptotic safety approach in 
Refs.~\cite{reutera,laut,reuterb,reuter3,litim,perc} 
to quantum gravity
may indeed provide a realization\footnote{The attendant 
choice of the scale $k\sim 1/t$ used in Refs.~\cite{reuter1,reuter2} 
was also proposed in Ref.~\cite{sola1}.} of the successful
inflationary model~\cite{guth,linde} of cosmology
without the need of the as yet unseen inflaton scalar field: the attendant UV fixed point solution
allows one to develop Planck scale cosmology that joins smoothly onto
the standard Friedmann-Walker-Robertson classical descriptions so
that then one arrives at a quantum mechanical 
solution to the horizon, flatness, entropy 
and scale free spectrum problems. In Ref.~\cite{bw2i}, we have shown
that, in the new
resummed theory~\cite{bw1,bw2,bw2a,bw2b,bw2c,bw2d,bw2e,bw2f,bw2g,bw2h} of quantum gravity, 
we recover the properties as used in Refs.~\cite{reuter1,reuter2} 
for the UV fixed point of quantum gravity with the
added results that we get ``first principles''
predictions for the fixed point values of
the respective dimensionless gravitational and cosmological constants
in their analysis. 
In what follows here, we carry the analysis one step further and arrive at an estimate
for the observed cosmological constant $\Lambda$ in the
context of the Planck scale cosmology of Refs.~\cite{reuter1,reuter2}.
We comment on the reliability of the result as well, as it will be seen
already to be relatively close to the observed value~\cite{cosm1,pdg2008}.
While we obviously do not want to overdo the closeness to the experimental value, we do want to argue that this again gives, at the least, some more credibility to the new resummed theory as well as to the methods in Refs.~\cite{reutera,laut,reuterb,reuter3,litim,perc,ambj}. More reflections on
the attendant implications of the latter credibility in the search for 
an experimentally testable union of the original ideas of Bohr and Einstein will be taken up
elsewhere~\cite{elswh}.\par
The discussion is organized as follows. We start by recapitulating 
the Planck scale cosmology presented phenomenologically
in Refs.~\cite{reuter1,reuter2}. This is done in the next section.
We then review our results in
Ref.~\cite{bw2i} for the dimensionless gravitational and cosmological constants
at the UV fixed point. In the course of this latter review, which is done in Section 3, we give
a new proof of the UV finiteness of the resummed quantum gravity
theory for the sake of completeness. In Section 4, we then
combine the Planck scale cosmology 
scenario in Refs.~\cite{reuter1,reuter2} with our results to estimate 
the observed value of 
the cosmological constant $\Lambda$. The Appendices contain relevant technical
details.
\par
\section{\bf Planck Scale Cosmology}
More precisely, we recall the Einstein-Hilbert 
theory
\begin{equation}
{\cal L}(x) = \frac{1}{2\kappa^2}\sqrt{-g}\left( R -2\Lambda\right)
\label{lgwrld1a}
\end{equation} 
where $R$ is the curvature scalar, $g$ is the determinant of the metric
of space-time $g_{\mu\nu}$, $\Lambda$ is the cosmological
constant and $\kappa=\sqrt{8\pi G_N}$ for Newton's constant
$G_N$. Using the phenomenological exact renormalization group
for the Wilsonian~\cite{kgw} coarse grained effective 
average action in field space, the authors in Ref.~\cite{reuter1,reuter2} 
have argued that
the attendant running Newton constant $G_N(k)$ and running 
cosmological constant
$\Lambda(k)$ approach UV fixed points as $k$ goes to infinity
in the deep Euclidean regime in the sense that 
$k^2G_N(k)\rightarrow g_*,\; \Lambda(k)\rightarrow \lambda_*k^2$
for $k\rightarrow \infty$ in the Euclidean regime.\par
The contact with cosmology then proceeds as follows. Using a phenomenological
connection between the momentum scale $k$ characterizing the coarseness
of the Wilsonian graininess of the average effective action and the
cosmological time $t$, the authors
in Refs.~\cite{reuter1,reuter2} show that the standard cosmological
equations admit of the following extension:
\begin{align}
(\frac{\dot{a}}{a})^2+\frac{K}{a^2}&=\frac{1}{3}\Lambda+\frac{8\pi}{3}G_N\rho\nonumber\\
\dot{\rho}+3(1+\omega)\frac{\dot{a}}{a}\rho&=0\nonumber\\
\dot{\Lambda}+8\pi\rho\dot{G_N}&=0\nonumber\\
G_N(t)&=G_N(k(t))\nonumber\\
\Lambda(t)&=\Lambda(k(t))
\label{coseqn1}
\end{align}
in a standard notation for the density $\rho$ and scale factor $a(t)$
with the Robertson-Walker metric representation as
\begin{equation}
ds^2=dt^2-a(t)^2\left(\frac{dr^2}{1-Kr^2}+r^2(d\theta^2+\sin^2\theta d\phi^2)\right)
\label{metric1}
\end{equation}
so that $K=0,1,-1$ correspond respectively to flat, spherical and
pseudo-spherical 3-spaces for constant time t.  Here, the equation of state
is taken as 
\begin{equation} 
p(t)=\omega \rho(t),
\end{equation}
where $p$ is the pressure. 
In Refs.~\cite{reuter1,reuter2}
the functional relationship between the respective
momentum scale $k$ and the cosmological time $t$ is determined
phenomenologically via
\begin{equation}
k(t)=\frac{\xi}{t}
\end{equation}
for some positive constant $\xi$ determined
from requirements on
physically observable predictions.\par
Using the UV fixed points as discussed above for $k^2G_N(k)\equiv g_*$ and
$\Lambda(k)/k^2\equiv \lambda_*$ obtained from their phenomenological, exact renormalization
group (asymptotic safety) 
analysis, the authors in Refs.~\cite{reuter1,reuter2}
show that the system in (\ref{coseqn1}) admits, for $K=0$,
a solution in the Planck regime where $0\le t\le t_{\text{class}}$, with
$t_{\text{class}}$ a ``few'' times the Planck time $t_{Pl}$, which joins
smoothly onto a solution in the classical regime, $t>t_{\text{class}}$,
which coincides with standard Friedmann-Robertson-Walker phenomenology
but with the horizon, flatness, scale free Harrison-Zeldovich spectrum,
and entropy\footnote{Here, we should note that, to solve the entropy problem, the authors in Ref.~\cite{reuter2} retain the general form of the requirement from Bianchi's identity so that the second and third relations in (\ref{coseqn1}) are combined to
$\dot{\rho}+3(1+\omega)\frac{\dot{a}}{a}\rho = -
\frac{\dot{\Lambda}+8\pi\rho\dot{G_N}}{8\pi G_N}$; we discuss this in more detail in Sect. 4.} problems all solved purely by Planck scale quantum physics.\par
While the dependencies of
the fixed-point results $g_*,\lambda_*$ on the cut-offs
used in the Wilsonian coarse-graining procedure, for example,
make the phenomenological nature of the analyses in Refs.~\cite{reuter1,reuter2} manifest, we note that 
the key properties of $g_*,\lambda_*$ used for these analyses 
are that the two UV limits are both positive and that the product 
$g_*\lambda_*$ is only mildly cut-off/threshold function dependent.
Here, we review the predictions in Refs.~\cite{bw2i} for these
UV limits as implied by resummed quantum gravity theory as presented in
~\cite{bw1,bw2,bw2a,bw2b,bw2c,bw2d,bw2e,bw2f,bw2g,bw2h}
and show how to use them to predict the current value of $\Lambda$.
In view of the lack of familiarity of the resummed quantum gravity theory,
we start the next section
with a review of its basic principles in the interest of making the discussion self-contained. 
\par
\section{\bf $g_*$ and $\lambda_*$ in Resummed Quantum  Gravity}
We start with the prediction for $g_*$, which we already presented in Refs.~\cite{bw1,bw2,bw2a,bw2b,bw2c,bw2d,bw2e,bw2f,bw2g,bw2h,bw2i}. Given that
the theory we use is not very familiar, we recapitulate
the main steps in the calculation in the interest of completeness.
\par
More specifically, as the graviton couples to a an elementary particle 
in the infrared regime which we shall
resum independently of the particle's spin, we may use a scalar
field to develop the required calculational framework. The
extension to spinning particles will then be straightforward.  
Thus, we start with the Lagrangian density for
the basic scalar-graviton system
which was considered by Feynman
in Refs.~\cite{rpf1,rpf2}:
\begin{equation}
\begin{split}
{\cal L}(x) &= -\frac{1}{2\kappa^2} R \sqrt{-g}
            + \frac{1}{2}\left(g^{\mu\nu}\partial_\mu\varphi\partial_\nu\varphi - m_o^2\varphi^2\right)\sqrt{-g}\\
            &= \quad \frac{1}{2}\left\{ h^{\mu\nu,\lambda}\bar h_{\mu\nu,\lambda} - 2\eta^{\mu\mu'}\eta^{\lambda\lambda'}\bar{h}_{\mu_\lambda,\lambda'}\eta^{\sigma\sigma'}\bar{h}_{\mu'\sigma,\sigma'} \right\}\\
            & \qquad + \frac{1}{2}\left\{\varphi_{,\mu}\varphi^{,\mu}-m_o^2\varphi^2 \right\} -\kappa {h}^{\mu\nu}\left[\overline{\varphi_{,\mu}\varphi_{,\nu}}+\frac{1}{2}m_o^2\varphi^2\eta_{\mu\nu}\right]\\
            & \quad - \kappa^2 \left[ \frac{1}{2}h_{\lambda\rho}\bar{h}^{\rho\lambda}\left( \varphi_{,\mu}\varphi^{,\mu} - m_o^2\varphi^2 \right) - 2\eta_{\rho\rho'}h^{\mu\rho}\bar{h}^{\rho'\nu}\varphi_{,\mu}\varphi_{,\nu}\right] + \cdots \\
\end{split}
\label{eq1-1}
\end{equation}
Here,
$\varphi(x)$ can be identified as the physical Higgs field as
our representative scalar field for matter,
$\varphi(x)_{,\mu}\equiv \partial_\mu\varphi(x)$,
and $g_{\mu\nu}(x)=\eta_{\mu\nu}+2\kappa h_{\mu\nu}(x)$
where we follow Feynman and expand about Minkowski space
so that $\eta_{\mu\nu}=diag\{1,-1,-1,-1\}$.
Following Feynman, we have introduced the notation
$\bar y_{\mu\nu}\equiv \frac{1}{2}\left(y_{\mu\nu}+y_{\nu\mu}-\eta_{\mu\nu}{y_\rho}^\rho\right)$ for any tensor $y_{\mu\nu}$\footnote{Our conventions for raising and lowering indices in the 
second line of (\ref{eq1-1}) are the same as those
in Ref.~\cite{rpf2}.}.
The bare(renormalized) mass of our otherwise free Higgs field is $m_o$($m$) 
and for the moment we set the small
observed~\cite{cosm1,pdg2008} value of the cosmological constant
to zero so that our quantum graviton, $h_{\mu\nu}$, has zero rest mass.
We return to the latter point, however, when we discuss phenomenology.
Feynman~\cite{rpf1,rpf2} has essentially worked out the Feynman rules for (\ref{eq1-1}), including the rule for the famous
Feynman-Faddeev-Popov~\cite{rpf1,ffp1a,ffp1b} ghost contribution needed 
for unitarity with the fixing of the gauge
(we use the gauge of Feynman in Ref.~\cite{rpf1},
$\partial^\mu \bar h_{\nu\mu}=0$),
so for this material we refer to Refs.~\cite{rpf1,rpf2}. 
Accordingly, we turn now directly to the quantum loop corrections
in the theory in (\ref{eq1-1}).
\par
\begin{figure}
\begin{center}
\epsfig{file=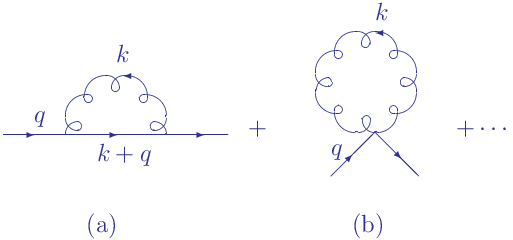,width=140mm}
\end{center}
\caption{\baselineskip=7mm     Graviton loop contributions to the
scalar propagator. $q$ is the 4-momentum of the scalar.}
\label{fig1}
\end{figure}

Referring to Fig.~\ref{fig1}, 
we have shown in Refs.~\cite{bw1,bw2,bw2a,bw2b,bw2c,bw2d,bw2e,bw2f,bw2g,bw2h} that the large virtual IR effects
in the respective loop integrals for 
the scalar propagator in quantum general relativity 
can be resummed to the {\em exact} result
\begin{equation}
\begin{split}
i\Delta'_F(k)&=\frac{i}{k^2-m^2-\Sigma_s(k)+i\epsilon}\cr
&=  \frac{ie^{B''_g(k)}}{k^2-m^2-\Sigma'_s+i\epsilon}\cr
&\equiv i\Delta'_F(k)|_{\text{resummed}}
\end{split}
\label{resum}
\end{equation}
for{\small ~~~($\Delta =k^2 - m^2$)
\begin{equation}
\begin{split} 
B''_g(k)&= -2i\kappa^2k^4\frac{\int d^4\ell}{16\pi^4}\frac{1}{\ell^2-\lambda^2+i\epsilon}\\
&\qquad\frac{1}{(\ell^2+2\ell k+\Delta +i\epsilon)^2}\\
&=\frac{\kappa^2|k^2|}{8\pi^2}\ln\left(\frac{m^2}{m^2+|k^2|}\right),       
\end{split}
\label{yfs1} 
\end{equation}}
where the latter form holds for the UV(deep Euclidean) regime, 
so that (\ref{resum}) 
falls faster than any power of $|k^2|$ -- by Wick rotation, the identification
$-|k^2|\equiv k^2$ in the deep Euclidean regime gives 
immediate analytic continuation to the result in the last line of (\ref{yfs1})
when the usual $-i\epsilon,\; \epsilon\downarrow 0,$ is appended to $m^2$. An analogous result~\cite{bw1} holds
for m=0; we show this in our Appendix 1 for completeness. Here, $-i\Sigma_s(k)$ is the 1PI scalar self-energy function
so that $i\Delta'_F(k)$ is the exact scalar propagator. As $\Sigma'_s$ starts in ${\cal O}(\kappa^2)$,
we may drop it in calculating one-loop effects. It follows that,
when the respective analogs of (\ref{resum}) are used for the
elementary particles, one-loop 
corrections are finite. It can be shown actually that the use of
our resummed propagators renders all quantum 
gravity loops UV finite~\cite{bw1,bw2,bw2a,bw2b,bw2c,bw2d,bw2e,bw2f,bw2g,bw2h}. We have called this representation
of the quantum theory of general relativity resummed quantum gravity (RQG).
\par
We stress that (\ref{resum}) is not limited to the regime where $k^2\cong m^2$
but is an identity that holds for all $k^2$. This is readily shown as follows.
If we invert both sides of (\ref{resum}) 
we get
\begin{equation}
\Delta_F^{-1}(k)-\Sigma_s(k)=(\Delta_F^{-1}(k)-\Sigma'_s(k))e^{-B''_g(k)}
\label{exact1}
\end{equation}
where the free inverse propagator is $\Delta_F^{-1}(k)=\Delta(k)+i\epsilon$.
We introduce here the loop expansions 
\begin{equation}
\Sigma_s(k)=\sum_{n=1}^{\infty}\Sigma_{s,n}(k),
\label{exact3a}
\end{equation}
\begin{equation}
\Sigma'_s(k)=\sum_{n=1}^{\infty}\Sigma'_{s,n}(k)
\label{exact3b}
\end{equation}
and we get, from elementary algebra, the exact relation
\begin{equation}
-\Sigma_{s,n}(k)=-\sum_{j=0}^{n}\Sigma'_{s,j}(k)\left(-B''_g(k)\right)^{n-j}/(n-j)!
\label{exact2}
\end{equation}
where we define for convenience $-\Sigma_{s,0}(k)=-\Sigma'_{s,0}(k)=\Delta_F^{-1}(k)$ and ${\cal A}_{s,n}$ is the n-loop contribution to ${\cal A}_s$. This proves that every Feynman diagram contribution to $\Sigma_s(k)$ corresponds to a unique contribution to $\Sigma'_s(k)$ to all orders in $\kappa^2/(4\pi)$ for all values of $k^2$. QED.\par

The key question is whether the terms which we have extracted from 
the Feynman series in (\ref{exact2}) 
were actually in that series. When we take the limit
that $k^2\rightarrow m^2$, the result is known to be valid from
the discussion in Ref.~\cite{wein-qft} where the same result for the
respective exponentiating
virtual infrared divergence in (\ref{yfs1}) is obtained. 
Indeed, one generally has to
introduce a regulator for the IR divergence and one shows that
the terms which diverge as the regulator vanishes exponentiate
in the factor $B''_g(k)$. When $k^2\ne m^2$, the IR divergence is regulated by
$\Delta(k)$, so that we can use $\Delta(k)$ as our IR regulator. We can then
isolate that part of the amplitude which diverges when $\Delta(k) \rightarrow 0$ when the UV divergences are themselves regulated, by n-dimensional methods~\cite{thft-vlt} for example, so that they remain finite in this limit.
At this point we stress the following: when we impose 
a gauge invariant regulator
for the UV regime, to any finite order in the loop expansion,
all UV divergences are regulated to finite results. If we then resum the IR
dominant terms in this the UV-regulated theory, that resummation is valid independent of whether or not
the theory is UV renormalizable, as the theory is finite 
order by order in the loop expansion in the UV
when the UV regulator is imposed independent of whether or not it is renormalizable. The latter issue arises only if we remove the UV regulator.
What we show now is that, after the IR resummation, the UV regulator can be removed and the UV regime remains finite order by order in the loop expansion
after the IR resummation.\par 
We call attention as well to the close analogy between our use of IR resummation in the presence of n-dimensional UV regularization to study the UV limit
of quantum gravity with the use of exact Wilsonian coarse graining
in Refs.~\cite{reutera,laut,reuterb,reuter3,litim,perc} to arrive at an effective average action 
for any given scale $k$ which has both an IR cut-off for momentum 
scales much smaller than $k$ and a UV cut-off for momentum scales much larger than $k$ so that the resulting field-space renormalization group
equation is well-defined even for a non-renormalizable theory like quantum gravity. In both cases the UV limit can be studied by taking the UV limit of the resulting non-perturbative solution and in both cases the same result obtains: a non-Gaussian UV fixed point is found, as we present below.\par
To show that (\ref{resum}) holds with $B''_g(k)$ given by the expression 
in (\ref{yfs1}), we proceed as follows. 
We represent the respective $m$-loop
contribution as defined above to the proper self-energy
contribution to the inverse propagator as 
\begin{equation}
i\Sigma_{s,m}(p)= \frac{1}{m!}\int\cdots\int \prod_{i=1}^{m}\frac{d^nk_i}{k_i^2-\lambda^2+i\epsilon}\rho_m(k_1,\cdots,k_m)
\label{am1}
\end{equation}
where $n$ is the analytically continued dimension
of space-time to regulate UV divergences and the function $\rho_m$ is symmetric under the interchange
of any two of the m virtual graviton n-momenta that are exchanged in
(\ref{am1}), by the Bose symmetry obeyed by the spin 2 gravitons
and the symmetry of the respective multiple integration volume.
Here is the point in the discussion where the power of exact rearrangement techniques such as those
in Ref.~\cite{yfs, yfs-jw} enters. For the case $m=1$, let $S''_g(k)\rho_0$ represent the leading contribution in the the limit $k\rightarrow 0$ to $\rho_1$.
We have
\begin{equation}
\rho_1(k) = S''_g(k)\rho_0+ \beta_1(k)
\label{indn0}
\end{equation}
where this equation is {\it exact} and serves to define $\beta_1$
if we specify $S''_g(k)$, the soft graviton emission factor, and recall
that 
\begin{equation}
\rho_0=i\Sigma_{s,0}(p)=-i\Delta_F(p)^{-1}.
\label{ir-1a}
\end{equation}
This can be determined from the Feynman rules for 
the Feynman~\cite{rpf1,rpf2,bw1} formulation of the 
scalar-graviton system in (\ref{eq1-1}) or
one can also use the off-shell extension of the formulas
in Ref.~\cite{wein-qft}. We get~\cite{bw1}
\begin{equation}
\begin{split}
S''_g(p,p,k) &= \frac{1}{(2\pi)^4}\frac{i\frac{1}{2}(\eta^{\mu\nu}\eta^{\bar\mu\bar\nu}+
\eta^{\mu\bar\nu}\eta^{\bar\mu\nu}-\eta^{\mu\bar\mu}\eta^{\nu\bar\nu})(-i\kappa p_{\bar\mu})(2ip_\mu)(-i\kappa{p'}_{\bar\nu})(2i{p'}_\nu)}{(k^2-2kp+\Delta+i\epsilon)(k^2-2kp'+\Delta'+i\epsilon)}{\Big|}_{p=p'}\\
           &= \frac{2i\kappa^2p^4}{16\pi^4}\frac{1}{(k^2-2kp+\Delta+i\epsilon)^2}
\end{split}
\label{reasf}
\end{equation}
where $\Delta'={p'}^2-m^2$. To see this, from Fig.~\ref{fig1}, note that the Feynman rules~\cite{rpf1,rpf2,bw1} give us the following result
\begin{equation}
\begin{split}
i\Sigma_{s,1}(p)&=\Big\{-\frac{\int d^nk}{(2\pi)^4}i v_3(p,p-k)_{\mu\bar{\mu}}\frac{i}{(p-k)^2-m^2+i\epsilon}i v_3(p-k,p')_{\nu\bar{\nu}}\\
&\qquad\qquad \frac{i\frac{1}{2}(\eta^{\mu\nu}\eta^{\bar\mu\bar\nu}+
\eta^{\mu\bar\nu}\eta^{\bar\mu\nu}-\eta^{\mu\bar\mu}\eta^{\nu\bar\nu})}{k^2-\lambda^2+i\epsilon}\\
&\qquad - \frac{\int d^nk}{2(2\pi)^4}i v_4(p,p')_{\mu\bar{\mu};\nu\bar{\nu}}\frac{i\frac{1}{2}(\eta^{\mu\nu}\eta^{\bar\mu\bar\nu} +
\eta^{\mu\bar\nu}\eta^{\bar\mu\nu} - \eta^{\mu\bar\mu}\eta^{\nu\bar\nu})}{k^2-\lambda^2+i\epsilon}\Big\}\Big{|}_{p=p'},
\end{split}
\label{ir-1}
\end{equation}
where we have defined from the Feynman rules the respective 3-point($h\varphi\varphi$ and 4-point($hh\varphi\varphi$) vertices 
\begin{equation}
\begin{split}
i v_3(p,p')_{\nu\bar{\nu}} &=-i\kappa\left(p_\nu{p'}_{\bar\nu}+p_{\bar\nu}{p'}_\nu-g_{\nu\bar{\nu}}(pp'-m^2)\right)\\
i v_4(p,p')_{\mu\bar{\mu};\nu\bar{\nu}} &= -4i\kappa^2[(pp'-m^2)(\eta_{\mu\bar{\nu}}\eta_{\bar{\mu}\nu}+\eta_{\bar{\mu}\bar{\nu}}\eta_{\mu\nu}-\eta_{\mu\bar{\mu}}\eta_{\nu\bar{\nu}}) \\
& \qquad\qquad - (p^{\mu'}{p'}^{\nu'}+p^{\nu'}{p'}^{\mu'})\{\eta_{\mu\mu'}(\eta_{\bar{\mu}\nu}\eta_{\nu'\bar{\nu}}+\eta_{\bar{\mu}\bar{\nu}}\eta_{\nu'\nu} - \eta_{\nu'\bar{\mu}}\eta_{\nu\bar{\nu}})\\
&\qquad\qquad + \eta_{\bar{\mu}\mu'}(\eta_{\mu\nu}\eta_{\nu'\bar{\nu}}+\eta_{\mu\bar{\nu}}\eta_{\nu'\nu} - \eta_{\nu'\mu}\eta_{\nu\bar{\nu}})\}]
\end{split}
\label{ir-2}
\end{equation}
using the standard conventions so that p is incoming and p' is outgoing for 
the scalar particle momenta at the respective vertices. In this way, we 
see that we may isolate the IR dominant part of $i\Sigma_1(p)$ by 
the separation
\begin{equation}
\begin{split}
\frac{1}{k^2-2kp+\Delta+i\epsilon}&=-\frac{\Delta}{(k^2-2kp+\Delta+i\epsilon)^2}+\frac{1}{k^2-2kp+i\epsilon}\\
&\qquad -\frac{2\Delta^2}{(k^2-2kp+\Delta+i\epsilon)^2(k^2-2kp+i\epsilon)}\\
&\qquad -\frac{\Delta^3}{(k^2-2kp+\Delta+i\epsilon)^2(k^2-2kp+i\epsilon)^2}\\
&\qquad +\sum_{n=2}^\infty(-1)^n\frac{\Delta^n}{(k^2-2kp+i\epsilon)^{n+1}}
\end{split}
\label{ir-3a}
\end{equation}
from which we can see that the first term on the RHS gives, upon insertion into (\ref{ir-1}), the IR-divergent
contribution for the coefficient of the lowest order inverse propagator
for the on-shell limit $\Delta\rightarrow 0$. The second term does not produce an IR-divergence and the remaining terms vanish faster than $\Delta$ in the on-shell limit so that they do not contribute to the field renormalization factor which we seek to isolate. In this way we get finally
\begin{equation}
\begin{split}
i\Sigma_1(p) &=
\Big\{-\frac{\int d^nk}{(2\pi)^4}(-2i\kappa p_\mu p_{\bar\mu}+i\delta{v_3(p,p-k)_{\mu\bar{\mu}}})(\frac{-i\Delta}{(k^2-2kp+\Delta+i\epsilon)^2}+i~R\Delta_F(k,p))\\
&\qquad\qquad (-2i\kappa {p}'_\nu {p'}_{\bar\nu}+i\delta{v_3(p'-k,p')_{\nu\bar{\nu}}})\frac{i\frac{1}{2}(\eta^{\mu\nu}\eta^{\bar\mu\bar\nu}+
\eta^{\mu\bar\nu}\eta^{\bar\mu\nu}-\eta^{\mu\bar\mu}\eta^{\nu\bar\nu})}{k^2-\lambda^2+i\epsilon}\\
&\qquad - \frac{\int d^nk}{2(2\pi)^4}i v_4(p,p')_{\mu\bar{\mu};\nu\bar{\nu}}\frac{i\frac{1}{2}(\eta^{\mu\nu}\eta^{\bar\mu\bar\nu} +
\eta^{\mu\bar\nu}\eta^{\bar\mu\nu} - \eta^{\mu\bar\mu}\eta^{\nu\bar\nu})}{k^2-\lambda^2+i\epsilon}\Big\}\Big{|}_{p=p'}\\
&\qquad = \Big\{\frac{\int d^nk}{(2\pi)^4}[(-i\kappa p_{\bar\mu})(2ip_\mu)\frac{-i\Delta}{(k^2-2kp+\Delta+i\epsilon)^2}(-i\kappa {p'}_{\bar\nu})(2i{p'}_\nu)\\
&\qquad\qquad \frac{i\frac{1}{2}(\eta^{\mu\nu}\eta^{\bar\mu\bar\nu}+
\eta^{\mu\bar\nu}\eta^{\bar\mu\nu}-\eta^{\mu\bar\mu}\eta^{\nu\bar\nu})}{k^2-\lambda^2+i\epsilon}+\frac{(2\pi)^4\beta_1(k)}{k^2-\lambda^2+i\epsilon}]\Big\}\Big{|}_{p=p'},
\end{split}
\label{ir-3b}
\end{equation}
which agrees with (\ref{indn0},\ref{ir-1a},\ref{reasf}) with
\begin{equation}
\begin{split}
R\Delta_F(k,p)&=\frac{1}{k^2-2kp+i\epsilon}
-\frac{2\Delta^2}{(k^2-2kp+\Delta+i\epsilon)^2(k^2-2kp+i\epsilon)}\\
&\qquad -\frac{\Delta^3}{(k^2-2kp+\Delta+i\epsilon)^2(k^2-2kp+i\epsilon)^2}\\
&\qquad +\sum_{n=2}^\infty(-1)^n\frac{\Delta^n}{(k^2-2kp+i\epsilon)^{n+1}},\\
i\delta{v_3(p,p-k)_{\mu\bar{\mu}}}&=i v_3(p,p-k)_{\mu\bar{\mu}}-\{-2i\kappa p_\mu p_{\bar\mu}\},\\
\beta_1(k)&= \Big\{-\frac{1}{(2\pi)^4}(-2i\kappa p_\mu p_{\bar\mu}+i\delta{v_3(p,p-k)_{\mu\bar{\mu}}})[\frac{-i\Delta}{(k^2-2kp+\Delta+i\epsilon)^2}\\
&\qquad\qquad +i~R\Delta_F(k,p)]
(i\delta{v_3(p'-k,p')_{\nu\bar{\nu}}})\{i\frac{1}{2}(\eta^{\mu\nu}\eta^{\bar\mu\bar\nu}+
\eta^{\mu\bar\nu}\eta^{\bar\mu\nu}-\eta^{\mu\bar\mu}\eta^{\nu\bar\nu})\}\\
&\qquad\qquad -\frac{1}{(2\pi)^4}(-2i\kappa p_\mu p_{\bar\mu}+i\delta{v_3(p,p-k)_{\mu\bar{\mu}}})(i~R\Delta_F(k,p))\\
&\qquad\qquad (-2i\kappa {p}'_\nu {p'}_{\bar\nu})\{i\frac{1}{2}(\eta^{\mu\nu}\eta^{\bar\mu\bar\nu}+
\eta^{\mu\bar\nu}\eta^{\bar\mu\nu}-\eta^{\mu\bar\mu}\eta^{\nu\bar\nu})\}\\
&\qquad\qquad -\frac{1}{(2\pi)^4}(i\delta{v_3(p,p-k)_{\mu\bar{\mu}}})(\frac{-i\Delta}{(k^2-2kp+\Delta+i\epsilon)^2})\\
&\qquad\qquad (-2i\kappa {p}'_\nu {p'}_{\bar\nu})\{i\frac{1}{2}(\eta^{\mu\nu}\eta^{\bar\mu\bar\nu}+
\eta^{\mu\bar\nu}\eta^{\bar\mu\nu}-\eta^{\mu\bar\mu}\eta^{\nu\bar\nu})\}\\
&\qquad\qquad - \frac{1}{2(2\pi)^4}i v_4(p,p')_{\mu\bar{\mu};\nu\bar{\nu}}\{i\frac{1}{2}(\eta^{\mu\nu}\eta^{\bar\mu\bar\nu} +
\eta^{\mu\bar\nu}\eta^{\bar\mu\nu} - \eta^{\mu\bar\mu}\eta^{\nu\bar\nu})\}\Big\}\Big{|}_{p=p'}.
\end{split}
\label{ir-4a}
\end{equation}
\par
One can see that the
result in (\ref{reasf}) differs from the corresponding result in QED in
Eq.(5.13) of Ref.~\cite{yfs} by the replacement of the electron charges
$e$ by the gravity charges $\kappa p_{\bar\mu},~\kappa{p'}_{\bar\nu}$
with the corresponding replacement of the photon propagator numerator
$-i\eta_{\mu\nu}$ by the graviton propagator numerator $i\frac{1}{2}(\eta^{\mu\nu}\eta^{\bar\mu\bar\nu}+
\eta^{\mu\bar\nu}\eta^{\bar\mu\nu}-\eta^{\mu\bar\mu}\eta^{\nu\bar\nu})$.
That the squared modulus of these gravity charges grows quadratically
in the deep Euclidean regime is what makes their effect therein in the quantum
theory of general relativity fundamentally different from the effect
of the QED charges in the deep Euclidean regime of QED, where the latter
charges are constants order-by-order in perturbation theory.\par

Indeed, proceeding recursively, we write
\begin{equation}
\rho_{m}(k_1,\cdots,k_m)=S''_g(k_m)\rho_{m-1}(k_1,\cdots,k_{m-1})+\beta^{(1)}_m(k_1,\cdots,k_{m-1};k_m)
\label{indn1}
\end{equation}
where here the notation indicates that the residual $\beta^{(1)}_m$ does not
contain the leading infrared contribution for $k_m$ that is given by the
first term on the RHS of (\ref{indn1})\footnote{ We stress that it may contain
in general other IR singular contributions.}. We iterate (\ref{indn1}) to get
\begin{equation}
\begin{split}
\rho_{m}(k_1,\cdots,k_m)&=S''_g(k_m)S''_g(k_{m-1})\rho_{m-2}(k_1,\cdots,k_{m-2})\\
&+S''_g(k_m)\beta^{(1)}_{m-1}(k_1,\cdots,k_{m-2};k_{m-1})\\
&+S''_g(k_{m-1})\beta^{(1)}_{m-1}(k_1,\cdots,k_{m-2};k_m)\\
&+\{-S''_g(k_{m-1})\beta^{(1)}_{m-1}(k_1,\cdots,k_{m-2};k_m)+\beta^{(1)}_m(k_1,\cdots,k_{m-1};k_m)\}
\end{split}
\label{indn2}
\end{equation}
The symmetry of $\rho_m$ implies that the quantity in curly brackets is
also symmetric in the interchange of $k_{m-1}$ and $k_m$. We indicate this
explicitly with the notation
\begin{equation}
\{-S''_g(k_{m-1})\beta^{(1)}_{m-1}(k_1,\cdots,k_{m-2};k_m)+\beta^{(1)}_m(k_1,\cdots,k_{m-1};k_m)\} = \beta^{(2)}_m(k_1,\cdots,k_{m-2};k_{m-1},k_m).
\label{indn3}
\end{equation}
\par

Repeated application of (\ref{indn1}) and use of the symmetry of
$\rho_m$ leads us finally to the {\it exact} result
\begin{equation}
\begin{split}
\rho_m(k_1,\cdots,k_m)&=S''_g(k_1)\cdots S''_g(k_m)\beta_0\\
&+\sum_{i=1}^{m}S''_g(k_1)\cdots S''_g(k_{i-1})S''_g(k_{i+1})\cdots S''_g(k_m)\beta_1(k_i) \\
&+ \cdots + \sum_{i=1}^{m}S''_g(k_i)\beta_{m-1}(k_1,\cdots,k_{i-1},k_{i+1},\cdots,k_m)+\beta_m(k_1,\cdots,k_m)
\end{split}
\label{indn3a}
\end{equation}
where the case m=1 has already been considered in (\ref{indn0})
with $\rho_0\equiv \beta_0$. Here, we defined as well $\beta^{(i)}_i \equiv \beta_i$.

We can use the symmetry of the residuals $\beta_i$ to re-write $\rho_m$
as
\begin{equation}
\rho_m(k_1,\cdots,k_m)=\sum_{perm}\sum_{r=0}^{m}\frac{1}{r!(m-r)!}\prod_{i=1}^rS''_g(k_i)\beta_{m-r}(k_{r+1},\cdots,k_m)
\label{indn4}
\end{equation}
so that we finally obtain, upon substitution into (\ref{am1}),
\begin{equation}
i\Sigma_{s,m}(p)=\sum_{r=0}^{m}\frac{1}{r!(m-r)!}\left(\int\frac{d^nk~S''_g(k)}{k^2-\lambda^2+i\epsilon}\right)^r\int\prod_{i=1}^{m-r}\frac{d^nk_i}{{k_i}^2-\lambda^2+i\epsilon}\beta_{m-r}(k_1,\cdots,k_{m-r}).
\label{indn5}
\end{equation}
With the definition
\begin{equation}
-B''_g(p)= \int\frac{d^nk~S''_g(k)}{k^2-\lambda^2+i\epsilon}
\label{indn6}
\end{equation}
and the identification
\begin{equation}
i\Sigma'_{s,r}(p) =\frac{1}{r!}\int\prod_{i=1}^{r}\frac{d^nk_i}{k_i^2-\lambda^2+i\epsilon}\beta_r(k_1,\cdots,k_r)
\label{indn7}
\end{equation}
we introduce the result (\ref{indn5}) into (\ref{exact1}) via (\ref{exact3a}) 
to get
\begin{equation}
\begin{split}
-i\left(\Delta_F(p)^{-1}-\Sigma_s(p)\right) &= i\sum_{m=0}^{\infty}\sum_{r=0}^m\Sigma'_{s,m-r}(p)\frac{(-B''_g(p))^r}{r!}\\
&= ie^{-B''_g(p)}\sum_{\ell=0}^{\infty}\Sigma'_{s,\ell}(p)\\
&= -ie^{-B''_g(p)}\left(\Delta_F(p)^{-1}-\sum_{\ell=1}^{\infty}\Sigma'_{s,\ell}(p)\right).
\end{split}
\label{indn8}
\end{equation}
In this way, our resummed {\it exact} result for the complete scalar propagator
in quantum general relativity is seen to be~\cite{bw1,bw2a,bw2b,bw2c}
\begin{equation}
\begin{split}
i\Delta'_F(p)&= \frac{ie^{B''_g(p)}}{(p^2-m^2-\Sigma'_s(p)+i\epsilon)}\cr
&\equiv i\Delta'_F(p)|_{\text{resummed}}\equiv i\Delta'_F(p)|_{\text{rsm}}
\end{split}
\label{indn9}
\end{equation}
where
\begin{equation}
\Sigma'_s(p)\equiv \sum_{\ell=1}^{\infty}\Sigma'_{s,\ell}(p).
\label{indn10}
\end{equation}
We have introduced the shorthand ``rsm'' for ``resummed'' in the last line
of (\ref{indn9}) for later convenience. 
\par
This result (\ref{indn10}) becomes identical to (\ref{resum}) when we take
the limit $n\rightarrow 4$ in it. In taking this limit, we note that 
$B''_g(k)$ is UV finite so that the limit exists without further ado.
As the IR limit of the coupling of the graviton to a particle
is well-known~\cite{wein-qft}
to independent of its spin, the entirely analogous result to (\ref{indn10}) 
holds for the propagators of all particles~\cite{bw1,bw2a,bw2b,bw2c} with corresponding exponent $B''_g(k)$ and the attendant IR-improved proper self-energy function. We note that in $\Sigma'_s(p)$ the limit $n\rightarrow 4$ 
can be taken if we represent it by its
IR-improved propagator expansion in which, to any finite order in
the loop expansion, the usual free Feynman propagator is replaced by its resummed version with the attendant IR-improved proper self-energy function,
$\Sigma'_s(p)$ or its graviton analog, set to zero on at least one internal 
line (per loop): for the scalar case this reads
\begin{equation}
i\Delta_F(p)|_{\text{resummed}} = \frac{ie^{B''_g(p)}}{(p^2-m^2+i\epsilon)}.
\end{equation}
with a corresponding result for the graviton case.
Standard resummation algebra then can be used to remove any double counting
effects to any finite order in the loop expansion, as $B''_g(k)$ is a UV finite
one-loop effect. Let us now see how one proves this last remark.\par
To this end, let $\Gamma^{\ell,m}(k_1,\ldots,k_\ell;k'_1,\ldots,k'_m)$
be the 1PI $\ell$-graviton, m-scalar proper vertex function, where we suppress
all Lorentz indices without loss of content.   
We follow Ref.~\cite{bj-d} and write $\Gamma^{\ell,m}(k_1,\ldots,k_\ell;k'_1,\ldots,k'_m)$ in terms of its skeleton expansion in which,
to any finite order in the respective loop expansion,
each graph ${\cal G}$ is mapped into a unique skeleton ${\cal S}$ in which all corrections to propagators and interaction vertices are removed. We then have the identification
\begin{equation}
\Gamma^{\ell,m}(k_1,\ldots,k_\ell;k'_1,\ldots,k'_m)=\sum_{\text{skeletons}\; {\cal S}}\Gamma^{{\cal S},\ell,m}(k_1,\ldots,k_\ell;k'_1,\ldots,k'_m;\Delta'_F,D'_F,\{\Gamma_j\},\kappa)
\label{bj-d1}
\end{equation} 
following the recipe in Ref.~\cite{bj-d} so that here
one uses the complete propagators, $\Delta'_F,D'_F$, for the scalar and the graviton on the lines of the skeleton and one uses the complete interaction
vertex foundations $\{\Gamma_j\}$ at each respective vertex in the skeleton to produce the exact, complete result for $\Gamma^{\ell,m}(k_1,\ldots,k_\ell;k'_1,\ldots,k'_m)$. In this representation, 
it is immediate how to obtain the attendant
$N$-th loop result accurate up to and including the $N$-th loop for  $\Gamma^{\ell,m}(k_1,\ldots,k_\ell;k'_1,\ldots,k'_m)$: one expands the propagators and complete interaction vertices to the appropriate
order, $\le N$ and retains all terms with $\le N$ loops in the sum on the RHS
of (\ref{bj-d1}). In the case of the exact scalar propagator, for example, 
we expand it as usual in each term in (\ref{bj-d1}),
\begin{equation}
i\Delta'_F(p)=i\Delta_F(p)+i\Delta_F(p)(-i\Sigma_s(p))i\Delta_F(p)+\cdots,
\end{equation} 
and we stop at the term with $N$-factors of $(-i\Sigma_s(k))$ each one of which we evaluate only to one loop order in this 
last term, with the attendant higher loop  evaluations in the 
terms with less than $N$ factors by the standard methodology. 
Inserting this result into (\ref{bj-d1})
with the analogous ones for the graviton propagator and the interaction vertices we isolate the result accurate up to and including the $N$-th loop by dropping all contributions that involve more than $N$-loops. This is the standard Feynman diagrammatic practice. Since we have the n-dimensional regulation of the UV divergences, the result we obtain this way is UV finite.\par
To improve it we substitute the resummed representation for the propagators, which we denote as we have above so that we have
\begin{equation}
\Gamma^{\ell,m}(k_1,\ldots,k_\ell;k'_1,\ldots,k'_m)=\sum_{\text{skeletons}\;{\cal S}}\Gamma^{{\cal S},\ell,m}(k_1,\ldots,k_\ell;k'_1,\ldots,k'_m;{\Delta'_F}|_{\text{rsm}},{D'_F}|_{\text{rsm}},\{\Gamma_j\},\kappa)
\label{bj-d2}
\end{equation} 
To obtain the IR-improved result correct up to an including the $N$-th 
IR-improved loop, we repeat the same steps as we did for the un-improved case: for example, we expand the scalar propagator as
\begin{equation}
\begin{split}
i\Delta'_F(p)&=\frac{ie^{B''_g(p)}}{(p^2-m^2-\Sigma'_s(p)+i\epsilon)}\cr
&=ie^{B''_g(p)}\left(\Delta_F(p)+\Delta_F(p)(-i\Sigma'_s(p))i\Delta_F(p)+\cdots\right)
\end{split}
\end{equation}
where we now stop the expansion at the term with $N$-factors of $(-i\Sigma'_s(p))$ in which each factor is only computed to one-loop order. We then introduce this IR-improved $N$-loop result for the scalar propagator and the analogous results for the graviton propagator and the interaction vertices
accurate as well to $N$ loops in the IR-improved loops
into the the RHS of (\ref{bj-d2}) and drop all terms with more than $N$ IR-improved loops. The result is now UV finite because the exponential
factor in the respective 
propagators render the integration in deep UV finite for any finite order in 
the interaction strength $\kappa$ because these exponential factors
fall faster than
any of the finite powers of the loop momenta that occur at finite orders in
$\kappa$ as given by the Feynman rules that follow from
Refs.~\cite{rpf1,rpf2} for (\ref{eq1-1}). 
\par
Finally, we observe that (\ref{exact2}) can be inverted to give as well the
identity
\begin{equation}
-\Sigma'_{s,n}(k)=-\sum_{j=0}^{n}\Sigma_{s,j}(k)\left(B''_g(k)\right)^{n-j}/(n-j)!
\label{exact2a}
\end{equation}
This allows us to employ the same result (\ref{bj-d2}) in calculating
the IR-improved self-energy so that it too is now UV finite with
our IR-improved resummation prescription. It follows that, to any finite order
in the IR-improved loop expansion, all $\Gamma^{\ell,m}(k_1,\ldots,k_\ell;k'_1,\ldots,k'_m)$ are UV finite. QED\par
As we have indicated above~\cite{bw1} 
and as Weinberg has shown in Ref.~\cite{wein-qft},
the IR limit of the coupling of the graviton to a particle is independent of its spin, so that we get the same exponential behavior in the resummed propagator for all particles
in the Standard Model. Indeed,
when we use our resummed propagator results, 
as extended to all the particles
in the SM Lagrangian and to the graviton itself, working now with the
complete theory
\begin{equation}
{\cal L}(x) = \frac{1}{2\kappa^2}\sqrt{-g} \left(R-2\Lambda\right)
            + \sqrt{-g} L^{\cal G}_{SM}(x)
\label{lgwrld1}
\end{equation}
where $L^{\cal G}_{SM}(x)$ is SM Lagrangian written in diffeomorphism
invariant form as explained in Refs.~\cite{bw1,bw2a}, we show in the Refs.~\cite{bw1,bw2,bw2a,bw2b,bw2c,bw2d,bw2e,bw2f,bw2g,bw2h} that the denominator for the propagation of transverse-traceless
modes of the graviton becomes ($M_{Pl}$ is the Planck mass)
\begin{equation}
q^2+\Sigma^T(q^2)+i\epsilon\cong q^2-q^4\frac{c_{2,eff}}{360\pi M_{Pl}^2},
\label{dengrvprp}
\end{equation}
where we have defined
\begin{equation}
\begin{split}
c_{2,eff}&=\sum_{\text{SM particles j}}n_jI_2(\lambda_c(j))\\
         &\cong 2.56\times 10^4
\end{split}
\label{c2eff}
\end{equation}
with $I_2$ defined~\cite{bw1,bw2,bw2a,bw2b,bw2c,bw2d,bw2e,bw2f,bw2g,bw2h}
by
\begin{equation}
I_2(\lambda_c) =\int^{\infty}_0dx x^3(1+x)^{-4-\lambda_c x}
\end{equation}
and with $\lambda_c(j)=\frac{2m_j^2}{\pi M_{Pl}^2}$ and~\cite{bw1,bw2,bw2a,bw2b,bw2c,bw2d,bw2e,bw2f,bw2g,bw2h}
$n_j$ equal to the number of effective degrees of particle $j$. 
For completeness, 
we repeat the derivation of (\ref{dengrvprp}) 
in our Appendix 2, using results from Appendix 3.
In arriving at the numerical value in (\ref{c2eff}), we take the SM
masses as follows: for the now presumed three massive neutrinos~\cite{neut,neuta},
we estimate a mass at $\sim 3$ eV; for
the remaining members
of the known three generations of Dirac fermions
$\{e,\mu,\tau,u,d,s,c,b,t\}$, we use~\cite{pdg2002,pdg2002a,pdg2004}
$m_e\cong 0.51$ MeV, $m_\mu \cong 0.106$ GeV, $m_\tau \cong 1.78$ GeV,
$m_u \cong 5.1$ MeV, $m_d \cong 8.9$ MeV, $m_s \cong 0.17$ GeV,
$m_c \cong 1.3$ GeV, $m_b \cong 4.5$ GeV and $m_t \cong 174$ GeV and for
the massive vector bosons $W^{\pm},~Z$ we use the masses
$M_W\cong 80.4$ GeV,~$M_Z\cong 91.19$ GeV, respectively.
We set the Higgs mass at $m_H\cong 126$GeV, in view of the
limit from LEP2~\cite{lewwg,lewwga} and recent observations from ATLAS and CMS~\cite{atlas-cms-2012}.
We note that (see the Appendix 1) when the
rest mass of particle $j$ is zero, such as it is for the photon and the gluon,
the value of $m_j$ turns-out to be
$\sqrt{2}$ times the gravitational infrared cut-off
mass~\cite{cosm1,pdg2008}, which is $m_g\cong 3.1\times 10^{-33}$eV.
We further note that, from the
exact one-loop analysis of Ref.\cite{tHvelt1}, it also follows (see Appendix 2)
that the value of $n_j$ for the graviton and its attendant ghost is $42$.
For $\lambda_c\rightarrow 0$, we have found the approximate representation
(see Appendix 3)
\begin{equation}
I_2(\lambda_c)\cong \ln\frac{1}{\lambda_c}-\ln\ln\frac{1}{\lambda_c}-\frac{\ln\ln\frac{1}{\lambda_c}}{\ln\frac{1}{\lambda_c}-\ln\ln\frac{1}{\lambda_c}}-\frac{11}{6}.
\end{equation} 
These results allow us to identify (we use $G_N$ for $G_N(0)$) 
\begin{equation}
G_N(k)=G_N/(1+\frac{c_{2,eff}k^2}{360\pi M_{Pl}^2})
\end{equation}
and to compute the UV limit $g_*$ as
\begin{equation}
g_*=\lim_{k^2\rightarrow \infty}k^2G_N(k^2)=\frac{360\pi}{c_{2,eff}}\cong 0.0442.
\end{equation}
We stress that this result has no threshold/cut-off effects in it.
It is a pure property of the known world.\par
Turning now to the prediction for $\lambda_*$, we use the Euler-Lagrange
equations to get Einstein's equation as 
\begin{equation}
G_{\mu\nu}+\Lambda g_{\mu\nu}=-\kappa^2 T_{\mu\nu}
\label{eineq1}
\end{equation}
in a standard notation where $G_{\mu\nu}=R_{\mu\nu}-\frac{1}{2}Rg_{\mu\nu}$,
$R_{\mu\nu}$ is the contracted Riemann tensor, and
$T_{\mu\nu}$ is the energy-momentum tensor. Working then with
the representation $g_{\mu\nu}=\eta_{\mu\nu}+2\kappa h_{\mu\nu}$
for the flat Minkowski metric $\eta_{\mu\nu}=\text{diag}(1,-1,-1,-1)$
we see that to isolate $\Lambda$ in Einstein's 
equation (\ref{eineq1}) we may evaluate
its VEV(vacuum expectation value of both sides). 
For any bosonic quantum field $\varphi$ we use
the point-splitting definition\footnote{We need to stress that this is a definition of convenience and is {\em not} a regularization because the integral which we calculate in (\ref{lambscalar}) below it is UV finite with exponential damping in the UV. The definition is robust, the direction of approach to the origin can be chosen arbitrarily, and when its vacuum expectation value is taken it may be replaced with the standard path integral Feynman rule for the tadpole loop that it most certainly is to give the same result.}  (here, :~~: denotes normal ordering as usual)
\begin{equation}
\begin{split}
\varphi(0)\varphi(0)&=\lim_{\epsilon\rightarrow 0}\varphi(\epsilon)\varphi(0)\cr
&=\lim_{\epsilon\rightarrow 0} T(\varphi(\epsilon)\varphi(0))\cr
&=\lim_{\epsilon\rightarrow 0}\{ :(\varphi(\epsilon)\varphi(0)): + <0|T(\varphi(\epsilon)\varphi(0))|0>\}\cr
\end{split}
\end{equation}
where the limit $\epsilon\equiv(\epsilon,\vec{0})\rightarrow (0,0,0,0)\equiv 0$
is taken from a time-like direction respectively. Thus, 
a scalar makes the contribution to $\Lambda$ given by\footnote{We note the
use here in the integrand of $2k_0^2$ rather than the $2(\vec{k}^2+m^2)$ in Ref.~\cite{bw2i}, to be
consistent with $\omega=-1$~\cite{zeld} for the vacuum stress-energy tensor.}
\begin{equation}
\begin{split}
\Lambda_s&=-8\pi G_N\frac{\int d^4k}{2(2\pi)^4}\frac{(2k_0^2)e^{-\lambda_c(k^2/(2m^2))\ln(k^2/m^2+1)}}{k^2+m^2}\cr
&\cong -8\pi G_N[\frac{1}{G_N^{2}64\rho^2}],\cr
\label{lambscalar}
\end{split}
\end{equation} 
where $\rho=\ln\frac{2}{\lambda_c}$ and we have used the calculus
of Refs.~\cite{bw1,bw2,bw2a,bw2b,bw2c,bw2d,bw2e,bw2f,bw2g,bw2h}
as recapitulated here in Appendices 2,3. 
The standard equal-time (anti-)commutation 
relations algebra realizations
then show that a Dirac fermion contributes $-4$ times $\Lambda_s$ to
$\Lambda$. The deep UV limit of $\Lambda$ then becomes, allowing $G_N(k)$
to run as we calculated,
\begin{equation}
\begin{split}
\Lambda(k) &\operatornamewithlimits{\longrightarrow}_{k^2\rightarrow \infty} k^2\lambda_*,\cr
\lambda_*&=-\frac{c_{2,eff}}{2880}\sum_{j}(-1)^{F_j}n_j/\rho_j^2\cr
&\cong 0.0817
\end{split}
\end{equation} 
where $F_j$ is the fermion number of $j$, $n_j$ is the effective
number of degrees of freedom of $j$ and $\rho_j=\rho(\lambda_c(m_j))$.
We see again that $\lambda_*$ is free of threshold/cut-off effects and is
a pure prediction of our known world -- $\lambda_*$ would vanish
in an exactly supersymmetric theory.\par
For reference, the UV fixed-point calculated here, 
$(g_*,\lambda_*)\cong (0.0442,0.0817)$, can be compared with the estimates
in Refs.~\cite{reuter1,reuter2}, 
which give $(g_*,\lambda_*)\approx (0.27,0.36)$. 
In making this comparison, one must keep in mind that 
the analysis in Refs.~\cite{reuter1,reuter2} did not include
the specific SM matter action and that there is definitely cut-off function
sensitivity to the results in the latter analyses. What is important
is that the qualitative results that $g_*$ and $\lambda_*$ are 
both positive and are less than 1 in size 
are true of our results as well.\par
For reference, we note that, if we restrict our resummed quantum gravity
calculations above for $g_*,\lambda_*$ to 
the pure gravity theory with no SM matter fields, we get the results
$$g_*=.0533,\;\lambda_*=-.000189.$$ We see that our results suggest
that there are still significant cut-off effects in the results 
used for $g_*,\;\lambda_*$ in Refs.~\cite{reuter1,reuter2}, which already
seem to include an effective matter contribution when viewed from
our resummed quantum gravity perspective, as an artifact of the obvious gauge and cut-off dependencies of the results. Indeed, 
from a purely quantum field theoretic point of view,
the cut-off action is 
\begin{equation}
\Delta_kS(h,C,\bar{C};\bar{g})=\frac{1}{2}<h,{\cal R}^{\text{grav}}_kh>+<\bar{C},{\cal R}^{\text{gh}}_kC>
\end{equation}
where $\bar{g}$ is the general background metric, which is the Minkowski space metric $\eta$ here, 
and $C,\bar{C}$ are the
ghost fields and the operators ${\cal R}^{\text{grav}}_k,\; {\cal R}^{\text{gh}}_k$ implement the course graining as they satisfy the limits 
\begin{equation}
\begin{split}
{\underset{p^2/k^2\rightarrow \infty}{\text{lim}}} {\cal R}_k &=0,\nonumber\\
{\underset{p^2/k^2\rightarrow 0}{\text{lim}}}{\cal R}_k&\rightarrow \mathfrak{Z}_k k^2,
\end{split}
\end{equation}
for some $\mathfrak{Z}_k$~\cite{laut}. Here, the inner product is that defined
in Ref.~\cite{laut} in its Eqs.(2.14,2.15,2.19).
The result is that the modes with $p\lesssim k$ have a shift of their vacuum energy
by the cut-off operator. There is therefore no disagreement in principle between
our gauge invariant results and the gauge dependent and cut-off dependent results in Refs.~\cite{laut}.
In other words, the graviton and ghost fields
at low scales compared to k have a mass added to them, so that their vacuum energies are shifted by a mass of order k. Evidently, this shows up as a positive
contribution to the cosmological constant and explains why the EFRG result
for $\lambda_*$ has a positive value in the regime of the gauge parameter
in Ref.~\cite{laut} where the UV fixed point is attractive.\par

\section{\bf An Estimate of $\Lambda$}

To see that the results here, taken together with those in Refs.~\cite{reuter1,reuter2}, allow us to estimate the value of $\Lambda$ today, we take the normal-ordered form of Einstein's equation 
\begin{equation}
:G_{\mu\nu}:+\Lambda :g_{\mu\nu}:=-\kappa^2 :T_{\mu\nu}: .
\label{eineq2}
\end{equation}
The coherent state representation of the thermal density matrix then gives
the Einstein equation in the form of thermally averaged quantities with
$\Lambda$ given by our result in (\ref{lambscalar}) summed over 
the degrees of freedom as specified above in lowest order. In Ref.~\cite{reuter2}, it is argued that the Planck scale cosmology description of inflation needs the transition time between the Planck regime and the classical Friedmann-Robertson-Walker(FRW) regime at $t_{tr}\sim 25 t_{Pl}$. (We comment below on the uncertainty of this choice of $t_{tr}$.)\footnote{The analysis in Ref.~\cite{reuter2} of their renormalization group
improved Einstein equations finds a set of solutions in which one has power law inflation in the UV regime and one switches
abruptly to the classical FRW solution with essentially zero cosmological constant at the transition time $t_{tr}$. In other words, 
the solution to the renormalization group improved Einstein equations at the transition time and later is very well approximated by non-running values of the gravitational and cosmological constant when one uses the FRW approximation. This also avoids issues of double counting of effects, for example. From our (\ref{eq-rho-lambda}) one sees that allowing the running to continue past $t_{tr}$ would not change our result for $\rho_\Lambda$ by very much at all, less than 8\%. We ignore effects of such size here.}
We thus introduce
\begin{equation}
\begin{split}
\rho_\Lambda(t_{tr})&\equiv\frac{\Lambda(t_{tr})}{8\pi G_N(t_{tr})}\cr
         &=\frac{-M_{Pl}^4(k_{tr})}{64}\sum_j\frac{(-1)^Fn_j}{\rho_j^2}
\end{split}
\label{eq-rho-lambda}
\end{equation}
and use the arguments in Refs.~\cite{branch-zap} ($t_{eq}$ is the time of matter-radiation equality) to get the 
first principles estimate, from the method of the operator field,
\begin{equation}
\begin{split}
\rho_\Lambda(t_0)&\cong \frac{-M_{Pl}^4(1+c_{2,eff}k_{tr}^2/(360\pi M_{Pl}^2))^2}{64}\sum_j\frac{(-1)^Fn_j}{\rho_j^2}\cr
          &\qquad\quad \times \frac{t_{tr}^2}{t_{eq}^2} \times (\frac{t_{eq}^{2/3}}{t_0^{2/3}})^3\cr
    &\cong \frac{-M_{Pl}^2(1.0362)^2(-9.194\times 10^{-3})}{64}\frac{(25)^2}{t_0^2}\cr
   &\cong (2.4\times 10^{-3}eV)^4.
\end{split}
\label{eq-rho-expt}
\end{equation}
where we take the age of the universe to be $t_0\cong 13.7\times 10^9$ yrs. 
In the latter estimate, the first factor in the second line comes from the period from
$t_{tr}$ to $t_{eq}$ which is radiation dominated and the second factor
comes from the period from $t_{eq}$ to $t_0$ which is matter dominated
\footnote{The method of the operator field forces the vacuum energies to follow the same scaling as the non-vacuum excitations.}.
This estimate should be compared with the experimental result~\cite{pdg2008}\footnote{See also Ref.~\cite{sola2} for an analysis that suggests 
a value for $\rho_\Lambda(t_0)$ that is qualitatively similar to this experimental result.} 
$\rho_\Lambda(t_0)|_{\text{expt}}\cong ((2.37\pm 0.05)\times 10^{-3}eV)^4$. 
\par
To sum up, in addition to our having put the  
Planck scale cosmology~\cite{reuter1,reuter2} on a
more rigorous basis, we believe our estimate 
of $\rho_\Lambda(t_0)$ represents some amount of progress in
the long effort to understand its observed value  
in quantum field theory. Evidently, the estimate is not a precision prediction,
as hitherto unseen degrees of freedom 
may exist and they have not been included, for example.\par
Indeed, we see that our result for the contribution to $\Lambda$ from a particle of rest mass $m$ scales as $1/\ln^2(2/\lambda_c(m))$ 
so that for masses $m<<M_{Pl}$
the larger the mass, the larger the contribution in magnitude. We note that the t, b, c, s, d, u,  $\tau$, $\mu$, e and the three neutrinos (together) contribute respectively 21.1\%, 17.6\%, 16.7\%, 
15.2\% , 13.5\%, 13.2\%, 5.63\%, 4.97\%, 4.01\% and 7.93\% of $\Lambda$ whereas the Higgs, W and Z 
bosons contribute -1.73\%, -5.10\% and -10.1\% of $\Lambda$ respectively.
The photon and the gluon, taken together, contribute -2.51\% of $\Lambda$, while the graviton contributes -0.277\%
thereof. Naively, such dependence on particle mass might appear to contradict the Appelquist-Carazzone decoupling theorem~\cite{ta-jc}, by which larger values of $m$ might be expected to be more suppressed. Two comments are in order. First, the decoupling theorem in Ref.~\cite{ta-jc} was only proved for renormalizable theories whereas the Einstein-Hilbert theory we 
deal with here is (power-countingly) nonrenormalizable. After we resum the theory, it is UV finite
with a characteristic scale of $\sim M_{Pl}$ for the scale beyond 
which the UV modes
are suppressed. Again, this is not the hypothesis of the Appelquist-Carazzone theorem. The key is the scale $M_{Pl}$. In the analyses presented above, we assume that $m/M_{Pl}<<1$ in deriving our results. For a quantity such as the 
integral on the RHS of the (\ref{lambscalar}) for $\Lambda_s$, 
which diverges like 4-powers of the cut-off without resummation and which
has a dependence on $M_{Pl}^4$ when we resum the theory, 
the remaining dependence on the
particle mass $m$ arises from the strength of the suppression of the modes beyond the characteristic scale $M_{Pl}$ and this is stronger for the smaller values of $m$ because they are farther away from $M_{Pl}$ which dominates the integral,
as we expect from the uncertainty principle. This phenomenon becomes even more
transparent if we consider masses $m>>M_{Pl}$, so that we are not subject
to effects of finite physical intrinsic scales. For two masses $m_1,\; m_2$
satisfying $m_i>> M_{Pl}$, we calculate that the contribution to $\Lambda_s$
scales as $m_iM_{Pl} $ so that we have the behavior one would expect from
summing the zero modes of a field of rest mass $m_i$ when the resummation 
causes the phase space integral to cut-off at a scale $\sim M_{Pl}$ yielding
the factor $-8\pi G_N(M_{Pl}^3m_i)$ since the vacuum energy density of the field is
given by (Here ${\cal H}$ is the usual free field Hamiltonian density.)
$$<0|{\cal H}|0>\sim \int^{M_{Pl}}\frac{ d^3 k}{(2\pi)^3}\frac{1}{2} \omega(k)= \int^{M_{Pl}}\frac{ d^3 k}{(2\pi)^3} \frac{1}{2}\sqrt{k^2+m_i^2}$$
where $\omega(k)$ is the usual frequency for mode $\vec{k}$ of the field and reduces
to $m_i$ when $ k^2<<m_i^2$. The larger mass makes a larger contribution because its zero modes are larger. This naturally raises the question of what would happen to our estimate if there would be a GUT theory at high scale?  We now comment on this.\par
In the current status of the standard GUT phenomenology, we know that the main viable
approaches involve susy GUT's because the standard
non-susy models have trouble to
match the value of $\sin^2\theta_W$ and have the three $SU_{2L}\times U_1\times SU(3)^c$ couplings~\cite{gsw, qcd} meet given their now precise values~\cite{pdg2008,siggi} at the scale $M_Z$, the rest mass of the $Z^0$ heavy gauge boson in the Glashow-Salam-Weinberg theory\cite{gsw}. To illustrate how a susy GUT might affect our estimate of $\Lambda$ we use
the susy SO(10) GUT model in Ref.~\cite{ravi-1} for definiteness.\par
In this model, the break-down of the GUT gauge symmetry to the 
low energy gauge symmetry occurs with an intermediate stage with gauge group
$SU_{2L}\times SU_{2R}\times U_1\times SU(3)^c$ where the final break-down to the Standard Model~\cite{gsw,qcd} gauge group, $SU_{2L}\times U_1\times SU(3)^c$, occurs at a scale $M_R\gtrsim 2TeV$ while the breakdown of global susy occurs at the (EW) scale $M_S$ which satisfies $M_R > M_S$. For our purposes
the key observation is that susy multiplets do not contribute to our formula
in (\ref{eq-rho-lambda}) when susy is not broken -- there is exact cancellation between fermions and bosons in a given degenerate susy multiplet. Thus only the the broken susy multiplets can contribute. In the model at hand, these are just the multiplets associated with the known SM particles and the extra Higgs multiplet required by susy in the MSSM~\cite{haber}.
In view of recent LHC results~\cite{lhc-susy}, we take for illustration the values $M_R\cong 4 M_S\sim 2.0{\text{TeV}}$ and set the following susy partner values:
\begin{equation}
\begin{split}
m_{\tilde{g}}&\cong 1.5(10){\text{TeV}}\\
m_{\tilde{G}}&\cong 1.5{\text{TeV}}\\
m_{\tilde{q}}&\cong 1.0{\text{TeV}}\\
m_{\tilde{\ell}}&\cong 0.5{\text{TeV}}\\
m_{\tilde{\chi}^0_i}&\cong\begin{cases} &0.4{\text{TeV}},\;i=1\\
                                        & 0.5{\text{TeV}},\; i=2,3,4
                    \end{cases}\\
m_{\tilde{\chi}^{\pm}_i}&\cong  0.5{\text{TeV}},\; i=1,2\\
m_{S}&= .5{\text{TeV}},\; S=A^0,\; H^{\pm},\; H_2,
\end{split}
\end{equation}  
where we use a standard notation for the susy partners of the known quarks($q\leftrightarrow \tilde{q}$), leptons($\ell\leftrightarrow \tilde{\ell}$) and gluons($G\leftrightarrow \tilde{G}$), and the EW gauge and Higgs bosons($\gamma,\; Z^0,\; W^{\pm},\;H,$\\
$A^0,\;H^{\pm},\;H_2  \leftrightarrow \tilde{\chi}$)  with the extra Higgs particles denoted as usual~\cite{haber} by $A^0$(pseudo-scalar), $H^{\pm}$(charged) and $H_2$(heavy scalar). $\tilde{g}$ is the gravitino, for which we show two examples of its mass for illustration. 
These particles then generate the extra contribution 
\begin{equation}
\begin{split}
\Delta W_{\rho,\text{GUT}}&=\sum_{j\in \{\text{MSSM low energy susy partners}\}}\frac{(-1)^Fn_j}{\rho_j^2}\\
          &\cong 1.13(1.12)\times 10^{-2}
\end{split} 
\end{equation}
to the factor $W_\rho\equiv \sum_j\frac{(-1)^Fn_j}{\rho_j^2}$ on the RHS of (\ref{eq-rho-lambda}) for the two respective values of $m_{\tilde g}$ called out by the parentheses. The corresponding values of $\rho_\Lambda$ are $-(1.67\times 10^{-3}\text{eV})^4(-(1.65\times 10^{-3}\text{eV})^4)$, respectively. The sign of these results would appear to put them in conflict with the positive observed value quoted above by many standard deviations, even when we allow for the considerable uncertainty in the various other factors multiplying $W_\rho$ in (\ref{eq-rho-lambda}), all of which are positive in our framework. This may be alleviated either by adding new particles to the model, approach (A), or by allowing a soft susy breaking mass term for the gravitino that resides near the GUT scale
$M_{GUT}$, which is $\sim 4\times 10^{16} GeV$ here~\cite{ravi-1}, approach (B). In approach (A), we double the number of quarks and leptons, but we invert
the mass hierarchy between susy partners, so that the new squarks and sleptons are lighter than the new quarks and leptons. This can work as long as
as we increase $M_R,\; M_S$ so that we have the new quarks and leptons
at $M_{\text{High}}\sim 3.4(3.3)\times 10^3\text{TeV}$ while leaving their partners at $M_{\text{Low}}\sim .5{\text{TeV}}$. For approach (B), the mass of the gravitino soft breaking term should be set to
$m_{\tilde{g}}\sim 2.3\times 10^{15}{\text{GeV}}$. More generally, our 
estimate in (\ref{eq-rho-expt}) can be used
as a constraint of general susy GUT models and we hope to explore such in more detail elsewhere. This admittedly 
limited discussion of susy GUT effects highlights what
one can expect for the impact on our estimate in (\ref{eq-rho-expt}) from higher mass
scale physics.\par 
Moreover, we need to stress that the value of $t_{tr}$ cannot be taken as precise, as we now elaborate. Specifically,
we are using for it the theory of Ref.~\cite{reuter2}. 
We can see that the solution to the renormalization group
improved Einstein equations
in Ref.~\cite{reuter2} relates $M_{Pl}\cong \xi H(t_{tr})\cong \alpha/t_{tr}$ where $\alpha =1/(2-2\Omega_\Lambda^*)$ with $\Omega_\Lambda^*$
 equal to the relative vacuum energy in the UV fixed point regime so that $\Omega_\Lambda^*\in (0,1)$. Here, $H$ is the Hubble parameter as usual and $\xi$
is of order unity and positive. For power law 
Planck scale inflation, we need $\alpha>1$, or  $\Omega_\Lambda^*> 1/2$. The authors in Ref.~\cite{reuter2} take as 'generic'
$\Omega_\Lambda^*=0.98$ which leads to $\alpha=25$ and in the solution to their renormalization group improved Einstein equations
to the $t_{tr}=\alpha t_{Pl}=25t_{Pl}$ that we have used here. 
Taking the difference between $\Omega_\Lambda^*$ and $1$ an order of magnitude smaller would amount to fine tuning,
so it is probably unreasonable. In addition, in order to match smoothly onto the FRW classical solution, $t_{tr}$ cannot
be too close to $t_{Pl}$, where the classical solution surely fails. Thus, we need $\alpha$ significantly larger than 1.
In other words, what the authors in Ref.~\cite{reuter2} have taken really does seem to be 'generic', as they put it. We feel $t_{tr}$ could be smaller
by a factor $\sim 3$ and could be larger by a similar factor and still be 'generic'. Even this error estimate alone would mean that our final result for $\rho_\Lambda$ is at least uncertain at the factor of $10$ level in the Bonanno and Reuter model. This should be taken in addition to the uncertainty associated
with the relation between the momentum scale $k$ and the cosmological time $t$ as we have indicated above for Ref.~\cite{reuter2},
where the estimates here realize this via Eqs.(2.2) and (5.1) in Ref.~\cite{reuter2}, $k(t)=\xi H(t)\cong \alpha/t$.\footnote{In our analysis, we work
on a flat background for our Fourier representations so that we have the 
usual Heisenberg connection between momentum space and position space -- our $k$ here is the not the same as the coarse graining scale $k$ in Ref.~\cite{reuter2}.} Given that we are switching from the Planck regime to the FRW regime,
there is uncertainty in $t_{tr}$ from both pieces of this last relation. Realistically, especially given the non-rigorousness of any argument based on fine tuning, we actually do not know the precise value of $t_{tr}$ at this point to better than a couple of orders of magnitude which translate to a {\em conservative} uncertainty at the level of $10^4$ on
our estimate of $\rho_\Lambda$. We caution the reader to keep this in mind.
\par
We discuss in closing three final important matters that we have not mentioned:(1), the effect of the various spontaneous symmetry vacuum energies on our 
$\rho_{\Lambda}$ estimate methodology as exhibited here; (2), the issue of the impact of our approach on big bang nucleosynthesis(BBN)~\cite{bbn}; and, (3), the covariance of theory in the presence of time dependent values of $\Lambda$ and of $G_N$. We consider these issues in turn, where we start with (1).\par 
From the standard methods we know for example that the energy of the broken vacuum for the EW case contributes an amount of order $M_W^4$ to $\rho_\Lambda$. If we consider the GUT 
symmetry breaking we expect an analogous contribution from spontaneous symmetry breaking of order $M_{GUT}^4$. When compared to the RHS of (\ref{eq-rho-lambda}), which is $\sim (-(1.0362)^2W_\rho/64)M_{Pl}^4\simeq \frac{10^{-2}}{64}M_{Pl}^4$, we see that adding these effects thereto would make relative changes in 
our results at the level of 
$\frac{64}{10^{-2}}\frac{M_W^4}{M_{Pl}^4}\cong 1\times 10^{-65} $ and $\frac{64}{10^{-2}}\frac{M_{GUT}^4}{M_{Pl}^4}\cong 7\times 10^{-7}$, respectively, where we use our value of $M_{GUT}$ given above in the latter 
evaluation for definiteness. We do ignore such small effects here.
\par
Concerning the impact, or the lack thereof, of our approach to $\Lambda$
on the phenomenology of big bang nucleosynthesis(BBN)~\cite{bbn}, we recall that the authors in Ref.~\cite{reuter2}
have already noted that when on passes from the Planck era to the FRW era,
a gauge transformation (from the attendant diffeomorphism invariance) is necessary to maintain consistency with
the solutions of the system (\ref{coseqn1})(or of its more general form as give below) at the boundary between the two regimes at the transition time $t_{tr}$. Requiring that the Hubble parameter be continuous at $t_{tr}$ 
the authors in Ref.~\cite{reuter2} arrive the gauge transformation on the
time for the FRW era relative to the Planck era \begin{equation}t\rightarrow t'=t-t_{as}\end{equation}
so that the continuity of the Hubble parameter at the boundary gives
\begin{equation}\frac{\alpha}{t_{tr}}=\frac{1}{2(t_{tr}-t_{as})}\end{equation}
when $a(t)\propto t^\alpha$ in the (sub-)Planck regime. This implies \begin{equation}t_{as}=(1-\frac{1}{2\alpha})t_{tr}.\end{equation} In our case , we have from Ref.~\cite{reuter2} the generic case $\alpha=25$, so that \begin{equation}t_{as}=0.98t_{tr}.\end{equation} Here, we have used the diffeomorphism invariance of the theory to choose another coordinate transformation for the FRW era, namely, \begin{equation}t\rightarrow t'=\gamma t\end{equation} 
as a part of a dilatation
where $\gamma$ now satisfies the boundary condition required for continuity of the Hubble parameter at $t_{tr}$:
\begin{equation}\frac{\alpha}{t_{tr}}=\frac{1}{2\gamma t_{tr}}\end{equation}  
so that \begin{equation}\gamma=\frac{1}{2\alpha}.\end{equation} The model in Ref.~\cite{reuter2} purports that, for $t>t_{tr}$, one has the time $t'$ and an effective FRW cosmology with such a small value of $\Lambda$ that it may be treated as zero. Here, we extend this by retaining $\Lambda\ne 0$ so that we may estimate its value. But, with our 
diffeomorphism transformation between the (sub-)Planck regime and the FRW regime, we can see that, at the time of BBN, the ratio of $\rho_\Lambda$ to
$\frac{3H^2}{8\pi G_N}$ is
\begin{equation} 
\begin{split} \Omega_\Lambda(t_{BBN}) &= \frac{M_{Pl}^2(1.0362)^29.194\times 10^{-3}(25)^2/(64 t_{BBN}^2)}{(3/(8\pi G_N))(1/(2\gamma t_{BBN})^2)}\cr &\cong \frac{\pi 10^{-2}}{24}\cr
&= 1.31\times 10^{-3}.\end{split}\label{bbneq1}\end{equation} Thus, at $t_{BBN}$ our $\rho_\Lambda$ is small enough that it has a negligible effect on the standard BBN phenomenology. We see that the uncertainty in the value of $\alpha$, which is the value of $t_{tr}$ in units of $\frac{1}{M_{Pl}},$ does not affect the estimate in (\ref{bbneq1}) because the factors of $\alpha^2=25^2$ cancel between the numerator and the denominator on the RHS in the first line of (\ref{bbneq1}). This is in contrast with our estimate of $\rho_\Lambda(t_0)$ in
(\ref{eq-rho-expt}) where the dependence on $\alpha^2=25^2$ is not cancelled, as we have discussed above.\par
Turning next to the issue of the covariance of the theory when $\Lambda$ and $G_N$ depend on time, we follow in Eqs.(\ref{coseqn1}) the corresponding realization of the improved Friedmann and Einstein equations as given in Eqs.(3.24) in Ref.~\cite{reuter1}. We note that the equations in (\ref{coseqn1}) should be 
compared to the more general
realization given in Eqs.(2.1) in Ref.~\cite{reuter2} -- we have effectively followed the latter realization in our discussions in this Section. The difference between the two realizations is the solution of the constraint following from Bianchi's identity:
\begin{equation}D^\nu\left(\Lambda g_{\nu\mu}+8\pi G_N T_{\nu\mu}\right)=0;\end{equation}
for, in (\ref{coseqn1}), this identity is solved for a covariantly conserved
$T_{\mu\nu}$ as well whereas in Eqs.(2.1) in Ref.~\cite{reuter2}, one has the modified conservation requirement, as we noted above,
\begin{equation}\dot{\rho}+3\frac{\dot{a}}{a}(1+\omega)\rho= -\frac{\dot{\Lambda}+8\pi\rho \dot{G}_N}{8\pi G_N}\end{equation}
to be compared with (\ref{coseqn1}) in which the RHS of this latter equation is set to zero. The phenomenology which we referenced from Ref.~\cite{reuter1} is qualitatively unchanged by the simplification in (\ref{coseqn1}) but of course the details of the that phenomenology, such as the (sub-)Planck era exponent for the time dependence of $a$, etc., are affected, as is the relation between $\dot{\Lambda}$ and
$\dot{G}_N$ in (\ref{coseqn1}). What we can say is that (\ref{coseqn1}) contains a special case of the more general realization of the Bianchi identity requirement when both $\Lambda$ and $G_N$ depend on time
whereas what we have done in this Section uses 
that more general realization. We should also note that only when $\dot{\Lambda}+8\pi\rho \dot{G}_N=0$ holds is covariant conservation of matter in the current universe guaranteed and that either the case with or the case without such guaranteed conservation is possible provided the attendant deviation is small. Detailed studies
of such deviation, including its maximum possible size, can be found in Refs.~\cite{bianref1,bianref2,bianref3}.\par
We want however to stress again that 
the model Planck scale cosmology of Bonanno and Reuter
which we use is just that, a model. More work needs to be done to remove 
from it the type of uncertainties which we just elaborated in our estimate of $\Lambda$.
We look forward, however, to additional possible checks from experiment 
with just this latter goal in mind. \par
\section*{Acknowledgments}
We thank Profs. L. Alvarez-Gaume and W. Hollik for the support and kind
hospitality of the CERN TH Division and the Werner-Heisenberg-Institut, MPI, Munich, respectively, where a part of this work was done.
\par
\section*{Note Added:}
Here, we point out for clarity that in computing $\Lambda$
in the Planck regime the assumption of $K=0$ is presumed 
as that is the only case for which the Bonanno-Reuter Planck scale
cosmology has been shown to allow a smooth connection from
the Planck regime for times near or earlier than the Planck time
to the semi-classical FRW regime for times after $t_{tr}$.
For $K=0$, by definition, equal time slices are flat 3-spaces, exactly
as we have employed in the vacuum states used to compute 
the zero-point energies that comprise $\Lambda$. Thus the results
in Sections 3 and 4 are fully self-consistent.
\section*{\bf Appendix 1: Evaluation of Gravitational Infrared Exponent}
In the text, we use several limits of the gravitational infrared
exponent $B''_g$ defined in (\ref{indn6}). Here, we present
these evaluations for completeness.\par

We have to consider
\begin{equation}
\begin{split}
-B''_g(p)&= \int\frac{d^4k~S''_g(k)}{k^2-\lambda^2+i\epsilon}\\
         &= \int\frac{d^4k}{(2\pi)^4(k^2-\lambda^2+i\epsilon)}\frac{i\frac{1}{2}(\eta^{\mu\nu}\eta^{\bar\mu\bar\nu}+
\eta^{\mu\bar\nu}\eta^{\bar\mu\nu}-\eta^{\mu\bar\mu}\eta^{\nu\bar\nu})(-i\kappa p_{\bar\mu})(2ip_\mu)(-i\kappa{p'}_{\bar\nu})(2i{p'}_\nu)}{(k^2-2kp+\Delta+i\epsilon)(k^2-2kp'+\Delta'+i\epsilon)}{\Big|}_{p=p'}\\
           &= \frac{2i\kappa^2p^4}{16\pi^4}\int\frac{d^4k}{(k^2-\lambda^2+i\epsilon)}\frac{1}{(k^2-2kp+\Delta+i\epsilon)^2}
\end{split}
\label{gexponent}
\end{equation}
where $\Delta=p^2-m^2$. The integral on the RHS of (\ref{gexponent}) is given by
\begin{equation}
\begin{split}
I&=\int\frac{d^4k}{(k^2-\lambda^2+i\epsilon)}\frac{1}{(k^2-2kp+\Delta+i\epsilon)^2}\nonumber\\
&= \frac{-i\pi^2}{p^2}\frac{1}{x_+-x_-}{\big[}x_+\ln(1-1/(\sqrt{2}x_+))-x_-\ln(1-1/(\sqrt{2}x_-)){\big]}
\end{split}
\end{equation}
with
\begin{equation}
x_\pm=\frac{1}{2\sqrt{2}}\left(\bar\Delta+\bar\lambda^2\pm((\bar\Delta+\bar\lambda^2)^2-4(\bar\lambda^2-i\bar\epsilon))^{\small 1/2}\right)
\end{equation}
for $\bar\Delta=1-m^2/p^2,~\bar\lambda^2=\lambda^2/p^2 \text{and}~\bar\epsilon=\epsilon/p^2$. In this way, we arrive at the results, for $p^2 <0$,
\begin{equation}
B''_g(p)
=\begin{cases}
&\frac{\kappa^2|p^2|}{8\pi^2}\ln\left(\frac{m^2}{m^2+|p^2|}\right),~~m\ne 0\\
&\frac{\kappa^2|p^2|}{8\pi^2}\ln\left(\frac{m_g^2}{m_g^2+|p^2|}\right),~~m=m_g=\lambda\\
&\frac{2\kappa^2|p^2|}{8\pi^2}\ln\left(\frac{m_g^2}{|p^2|}\right),~~m=0,~m_g=\lambda
\end{cases}
\label{irexpnt1}
\end{equation}
where we have made more explicit the presence of the
observed small mass, $m_g$, of the graviton. When m=0 and one wants to use dimensional regularization for the IR regime instead of $m_g$, we normalize the 
propagator at a Euclidean point $k^2=-\mu^2$ and use standard factorization
arguments~\cite{qcdfactorzna,qcdfactorznb,qcdfactorznc,qcdfactorznd,qcdfactorzne} to take the factorized result for $B''_g$ from
(\ref{irexpnt1}) as
\begin{equation}
B''_g(p)|_{\text{factorized}}=\frac{2\kappa^2|p^2|}{8\pi^2}\ln\left(\frac{|\mu^2|}{|p^2|}\right),~~m=0,~m_g=0.
\label{irexpnt2}
\end{equation}
In physical applications, such mass singularities are absorbed by the 
definition of the initial state ``parton'' densities and/or are canceled 
by the KLN theorem in the final state; we do not exponentiate them in 
the exactly massless case.
\par
We stress that the standard analytic properties of the 1PI 2pt functions
obtain here, as we use standard Feynman rules. Wick rotation changes the Minkowski space Feynman loop integral $\int d^4k$ with $k=(k^0,k^1,k^2,k^3)$ for real $k^j$ and $k^2={k^0}^2-{k^1}^2-{k^2}^2-{k^3}^2$ into the integral $i\int d^4k_E$ with $k=(ik^0,k^1,k^2,k^3)$ and $k^2=-{k^0}^2-{k^1}^2-{k^2}^2-{k^3}^2\equiv -k_E^2$ with $k_E$ the Euclidean 4-vector $k_E=(k^0,k^1,k^2,k^3)$ with metric $\delta_{\mu\nu}=diag(1,1,1,1)$. Thus our results rigorously correspond to $|p^2|=-p^2$ in (\ref{irexpnt1}),~(\ref{irexpnt2}) with $m^2$ replaced with $m^2-i\epsilon$,
with $\epsilon\downarrow 0$, following Feynman, for $p^2 <0$;  by Wick rotation this is the regime relevant to the UV behavior of the Feynman loop integral.
Standard complex variables theory then uniquely specifies our exponent
for any value of $p^2$. 
\par

\section*{\bf Appendix 2: Graviton Inverse Propagator}

\begin{figure}
\begin{center}
\epsfig{file=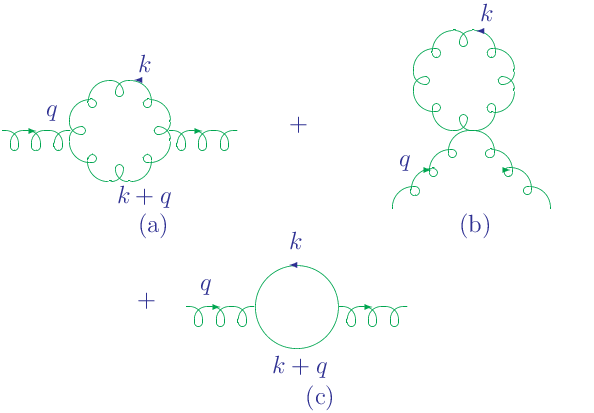,width=77mm,height=38mm}
\end{center}
\caption{\baselineskip=7mm  The graviton((a),(b)) and its ghost((c)) one-loop contributions to the graviton propagator. $q$ is the 4-momentum of the graviton.}
\label{fig2}
\end{figure}

\begin{figure}
\begin{center}
\epsfig{file=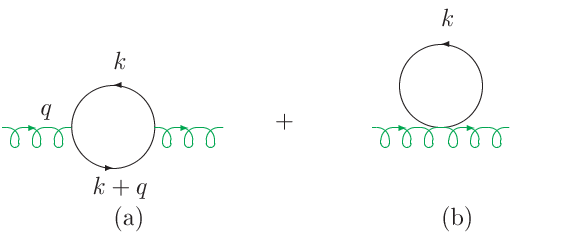,width=77mm,height=38mm}
\end{center}
\caption{\baselineskip=7mm  The scalar one-loop contribution to the
graviton propagator. $q$ is the 4-momentum of the graviton.}
\label{fig3}
\end{figure}

To obtain the result
in (\ref{dengrvprp}) we first consider~\cite{bw1} 
the diagrams in Figs.~\ref{fig2} and \ref{fig3}.
These graphs have a superficial degree of divergence
in the UV of +4 and are a test of our methods
because, in the usual treatment
of the theory, they generate
a UV divergence in the respective 1PI 2-point function
for the coefficient of $q^4$ which can not be
removed by the standard field and mass renormalizations.
\par

For example, consider the graph in Fig.~3a. When we use our resummed
propagators, we get
(here, $k\rightarrow (ik^0,\vec k)$ by Wick rotation, and we work in the
transverse-traceless space){\small
\begin{equation}
\begin{split}
i\Sigma(q)^{1a}_{\bar\mu\bar\nu;\mu\nu}&=i\kappa^2\frac{\int d^4k}{2(2\pi)^4}
\frac{\left(k'_{\bar\mu}k_{\bar\nu}+k'_{\bar\nu}k_{\bar\mu}\right)e^{\frac{\kappa^2|{k'}^2|}{8\pi^2}\ln\left(\frac{m^2}{m^2+|{k'}^2|}\right)}}
{\left({k'}^2-m^2+i\epsilon\right)}\\
   &\qquad   \frac{
\left(k'_{\mu}k_{\nu}+k'_{\nu}k_{\mu}\right)e^{\frac{\kappa^2|k^2|}{8\pi^2}\ln\left(\frac{m^2}{m^2+|k^2|}\right)}}
{\left(k^2-m^2+i\epsilon\right)}.
\end{split}
\label{eq2p}
\end{equation}}\noindent
We see explicitly that the exponential damping in the deep
Euclidean regime has rendered the graph in Fig.~3a finite in the UV.
For the same reason, all of the graphs in Figs. 2 and 3 are UV finite 
when we use
our respective resummed propagators to compute them.\par

To evaluate the effect of the corrections in Figs.~\ref{fig2} and \ref{fig3}
on the graviton propagator, we continue to work in the
transverse, traceless space and isolate the effects from
Figs.~\ref{fig2} and \ref{fig3} on the coefficient of the
$q^4$ in the graviton propagator denominator,
\begin{equation}
q^2 +\frac{1}{2}q^4\Sigma^{T(2)}+i\epsilon,
\label{prop}
\end{equation}
so that we need to evaluate the transverse, traceless self-energy
function $\Sigma^T(q^2)$ that
follows from (\ref{eq2p}) for Fig. 3a and its analogs for Figs. 3b
and 2 by the standard methods. Here, we work in the
expectation that, in consequence to the newly UV finite calculated
quantum loop effects in Figs. 2 and 3, the Fourier
transform of the graviton propagator that enters Newton's law,
our ultimate goal here, will receive support from
from $|q|^2<<M_{Pl}^2$. We will therefore work
in the limit that $q^2/M_{Pl}^2$ is relatively small,~$\lesssim .1$,
for example~\footnote{This regime is for numerical convenience only, as it allows us to work with a simple quadratic equation in $q^2$ in
determining the Fourier transform of the graviton propagator below. It is justified because the pole position which we find at non-zero $q^2$ satisfies it. There is no problem of principle to treat the exact result, and it will appear elsewhere.}. This will allow us to see the dominant effects
of our new finite quantum loop effects. In other words, we will
work to $\sim 10\%$ (leading-log) accuracy in what follows. See Appendix 2
for more discussion on this point.\par
 
First let us dispense with the contributions from Figs. 2b
and Fig. 3b. These are independent of $q^2$ so that
we use a mass counter-term to remove them and set the graviton mass
to $0$. 
Following the suggestion of Feynman in Ref.~\cite{rpf2}, 
we will change this to a small non-zero value below
to take into account the recently established small value of the
cosmological constant~\cite{cosm1,pdg2008}. See also the discussion in 
Ref.~\cite{novelloa,novellob,novelloc,novellod} where it is 
shown that the quantum fluctuations in the exact de Sitter metric implied by 
the non-zero cosmological constant correspond in general to a 
mass for the graviton. 
Here, as we expand about a flat background, we take this effect into account 
as a small infrared regulator for the graviton. The deviations from flat space 
in the deep Euclidean region that we study due to the observed value
of the cosmological constant are at the level of $e^{10^{-61}}-1$! This is
safely well beyond the accuracy of our methods.\par

Returning to Fig.~3a, when we project onto the transverse, traceless
space, that is to say, the graviton helicity space $\{e^{\mu\nu}(\pm2)=\epsilon^\mu_\pm\epsilon^\nu_{\pm},\, \text{where}\, \epsilon^\nu_{\pm}=\pm(\hat{x}\pm i\hat{y})/\sqrt{2}\, \text{when}\, \\ \hat{x},~\hat{y}\, \text{are purely space-like and} (\vec{\hat{x}},\vec{\hat{y}},\vec{q}/|\vec{q}|)\, \text{form a right-handed coordinate
basis}\}$,  we get (see the Appendix 3) the result
\begin{equation}
i\Sigma^T(q^2)_{3a}=\frac{-i\kappa^2m^4}{96\pi^2}\int^1_0d\alpha\int^{\infty}_0 dx\frac{x^3(2(x+1)\bar{d}+\bar{d}^2)}{(x+1)^2(x+1+\bar{d})^2}(1+x)^{-\lambda_c x}
\label{eq3p}
\end{equation}
where $\lambda_c=\frac{2m^2}{\pi M_{Pl}^2}$,~$\bar{d}=\alpha(1-\alpha)\vec{q}^2/m^2$ so that
we have made the substitution $x=k^2$ and imposed the mass counter-term
as we noted. We have taken for definiteness $q=(0,\vec{q})$. We also
use $q=|\vec{q}|$ when there is no chance for confusion.
We are evaluating (\ref{eq3p}) in the deep UV where $m^2/q^2<< 1$
and where $q^2/M_{Pl}^2\lesssim 0.1$ -- see footnote 8. Accordingly,
we get
\begin{equation}
i\Sigma^T(q^2)_{3a}=\frac{-i\kappa^2}{96\pi^2}\left(\frac{|\vec{q}|^2m^2c_1}{3}+\frac{|\vec{q}|^4c_2}{30}\right)
\end{equation}
where
\begin{equation}
\begin{split}
c_1=I_1(\lambda_c)&=\int^{\infty}_0dx x^3(1+x)^{-3-\lambda_c x}\\
c_2=I_2(\lambda_c)&=\int^{\infty}_0dx x^3(1+x)^{-4-\lambda_c x}.
\end{split}
\end{equation}
Using the usual field renormalization, we see that
Fig. 3a makes the contribution
\begin{equation}
i\tilde\Sigma^T(q^2)_{3a}\cong \frac{-i\kappa^2|\vec{q}|^4c_2}{2880\pi^2}
\end{equation}
to the transverse traceless graviton proper self-energy function.\par

Turning now to Figs. 2, the pure gravity loops, we use a contact
between our work and that of Refs.~\cite{tHvelt1}.
In Refs.~\cite{tHvelt1}, the entire set of one-loop divergences
has been computed for the theory in (\ref{eq1-1}).
The basic observation is the following. As we work only to the leading
logarithmic accuracy in $\ln \lambda_c$, it is sufficient to identify
the correspondence between the divergences as calculated in the n-dimensional
regularization scheme in Ref.~\cite{tHvelt1} and as they would occur
when $\lambda_c\rightarrow 0$. This we do by comparing our result
for (\ref{eq3p}) when $q^2\rightarrow 0$ with the corresponding result in
Ref.~\cite{tHvelt1} for the same theory. In this way we see that
we have the correspondence
\begin{equation}
-\ln\lambda_c \leftrightarrow \frac{1}{2-n/2}.
\label{uv1}
\end{equation}
This allows us to read-off the leading log result for the
pure gravity loops directly from the results in Ref.~\cite{tHvelt1}.
Since $-\ln\lambda_c = \ln {M_{Pl}}^2-\ln m^2 -\ln\frac{2}{\pi}$,
we see that our exponentiated propagators have cut-off our UV divergences
at the scale $\sim M_{Pl}$ and the correspondence in (\ref{uv1}) shows the usual relation between the effective UV cut-off scale and the pole in $(2-n/2)$
in dimensional regularization. Note as well that, if the small cosmological constant\cite{cosm1,pdg2008} is set to zero\footnote{For the reader unfamiliar with Feynman's original observation~\cite{rpf2} that, in his approach to QGR, one of the main effects
of the cosmological constant is to give the quantum graviton field $h_{\mu\nu}$ a mass, we recall Einstein's equation $R_{\mu\nu}-\frac{1}{2}g_{\mu\nu}R+\Lambda g_{\mu\nu}=-\kappa^2T_{\mu\nu}$, with $R_{\mu\nu}$ and $T_{\mu\nu}$ the respective Ricci and energy-momentum tensors. For $g_{\mu\nu}=\eta_{\mu\nu}+2\kappa h_{\mu\nu}$, we get $R_{\mu\nu}= \kappa r_{\mu\nu}+{\cal O}(\kappa^2)$, with $r_{\mu\nu}=\Box h_{\mu\nu}-\partial_\alpha\partial_\mu h_\nu^\alpha-\partial_\alpha\partial_\nu h_\mu^\alpha+\partial_\mu\partial_\nu h_\alpha^\alpha$ so that, absorbing the $\Lambda \eta_{\mu\nu}$ term into the normal ordering constant-$\eta_{\mu\nu}$ term in $T_{\mu\nu}$, we get the result $r_{\mu\nu}-\frac{1}{2}\eta_{\mu\nu}r_\alpha^\alpha +2\Lambda h_{\mu\nu}=\kappa T'_{\mu\nu}$ 
where here $T'_{\mu\nu}$ is now the normal ordered
energy-momentum tensor, including the contribution from the graviton itself.
This result shows that the field $h_{\mu\nu}$, as already noted by Feynman~\cite{rpf2}, now has mass-squared $2\Lambda$ working to leading order in $\Lambda$.
We treat this as an IR regulator mass for a massless spin 2 field
in Minkowski space over the Planck scale distances with which we work. Indeed, the non-zero value of $\Lambda$ means the background metric should be of de Sitter type and this avoids the problems noted in Refs.~\cite{v-d-velt-zaka,v-d-velt-zakb} associated with a graviton mass different from
zero in Minkowski space, as we explained further in the text above.}
, the graviton is then exactly massless and we normalize its
propagator at a Euclidean point $p^2=-\mu^2$ as is standard for massless non-Abelian gauge theories for example. It follows that for the graviton case and for all other cases where $m=0$, 
as we explain in Appendix 1 (see (\ref{irexpnt2})), the
mass $m$ in (\ref{uv1}) is replaced with $m=\mu$ -- there
is no zero mass divergence in the case that the mass of the 
respective particle is zero. The UV correspondence is the same in both the
$m\ne 0$ and $m=0$ cases.  
\par
Specifically, the result in Ref.~\cite{tHvelt1}, when interpreted as we
have just explained, is that the pure gravity
loops give a factor of 42 times the scalar loops for the coefficient
$a_2$ above when we work in the regime where $|q^2|$ is relatively
small compared to $M_{Pl}^2$. Here, we again take into account the recent evidence for a non-zero cosmological constant~\cite{cosm1,pdg2008}, which can be seen to provide the small non-zero rest mass for the graviton, $m_g\cong 3.1\times 10^{-33}$eV, which serves as an IR regulator for the graviton. This is the value of rest mass in $\lambda_c$ which should be used for pure gravitational
loops -- see footnote~9 for more discussion on this point relevant to Refs.~\cite{v-d-velt-zaka,v-d-velt-zakb}. See the Appendix 1 for the derivation of the
corresponding infrared exponents.
\par
We note that, for $\lambda_c=0$, the constant $c_2$ is infinite
and, as we have already imposed both the mass and field renormalization
counter-terms, there would be no physical parameter into which
that infinity could be absorbed: this is just another manifestation
that QGR, without our resummation, is a non-renormalizable theory.\par
\par
Using the universality of the coupling of the graviton when
the momentum transfer scale is relatively small compared to $M_{Pl}$,
we can extend the result for the scalar field above to the remaining
known particles in the Standard Model by counting the number of physical
degrees of freedom for each such particle and replacing the mass of the
scalar with the respective mass of that particle. For a massive fermion
we get a factor of 4 relative to the scalar result with the appropriate
change in the mass parameter from $m$ to $m_f$, the mass of that fermion,
for a massive vector,
we get a factor of 3 relative to the scalar result, with the corresponding
change in the mass from $m$ to $m_V$, the mass of that vector, etc.
In this way, we arrive at the result that the denominator of the
graviton propagator becomes, in the Standard Model,
\begin{equation}
q^2+\Sigma^T(q^2)+i\epsilon\cong q^2-q^4\frac{c_{2,eff}}{360\pi M_{Pl}^2},
\label{dengrvprpa}
\end{equation}
where we have defined
\begin{equation}
\begin{split}
c_{2,eff}&=\sum_{\text{SM particles j}}n_jI_2(\lambda_c(j))\\
         &\cong 2.56\times 10^4
\end{split}
\label{c2effa}
\end{equation}
with $I_2$ defined above and with $\lambda_c(j)=\frac{2m_j^2}{\pi M_{Pl}^2}$ and~\cite{bw2b}
$n_j$ equal to the number of effective degrees of particle $j$ as
already illustrated. The
values for Standard Model masses used
in arriving at the numerical value for $c_{2,eff}$ in (\ref{c2effa}) are 
explained in the text. We also note that (see Appendix 3) 
for $\lambda_c\rightarrow 0$, we have found the approximate representation
\begin{equation}
I_2(\lambda_c)\cong \ln\frac{1}{\lambda_c}-\ln\ln\frac{1}{\lambda_c}-\frac{\ln\ln\frac{1}{\lambda_c}}{\ln\frac{1}{\lambda_c}-\ln\ln\frac{1}{\lambda_c}}-\frac{11}{6}.
\label{I2eq-appx}
\end{equation}
\par
The results (\ref{dengrvprpa}), (\ref{c2effa}) and (\ref{I2eq-appx}) have 
been used in the text.
\par

\section*{\bf Appendix 3: Evaluation of Gravitationally Regulated Loop Integrals}
In this section we present the derivation of the representations
which we have used in the text in evaluating the gravitationally
regulated loop integrals in Figs.~\ref{fig2},\ref{fig3}.\par
Considering the integrals in Fig.~\ref{fig3} to show the methods, we need the
result for
\begin{equation}
\begin{split}
{\cal I}_{\bar\mu\bar\nu;\mu\nu}&=i\frac{\int d^4k}{(2\pi)^4}
\frac{\left(k'_{\bar\mu}k_{\bar\nu}+k'_{\bar\nu}k_{\bar\mu}\right)e^{\frac{\kappa^2|{k'}^2|}{8\pi^2}\ln\left(\frac{m^2}{m^2+|{k'}^2|}\right)}}
{\left({k'}^2-m^2+i\epsilon\right)}\\
   &\qquad   \frac{
\left(k'_{\mu}k_{\nu}+k'_{\nu}k_{\mu}\right)e^{\frac{\kappa^2|k^2|}{8\pi^2}\ln\left(\frac{m^2}{m^2+|k^2|}\right)}}
{\left(k^2-m^2+i\epsilon\right)}.
\end{split}
\label{a21}
\end{equation}
In the limit that $|q^2|<<M_{Pl}^2$, standard symmetric integration methods
give us, for the transverse parts,
\begin{equation}
{\cal I}_{\bar\mu\bar\nu;\mu\nu}=\frac{i\pi^2}{12}\{g_{\bar\mu\bar\nu}g_{\mu\nu}+permutations\}I_0
\end{equation}
where we have
\begin{equation}
I_0\cong\frac{\int_0^1d\alpha\int_0^\infty dkk^3}{(2\pi)^4}\frac{k^4e^{\lambda_c(k^2/m^2)\ln(m^2/(m^2+k^2))}}{[k^2+m^2+|q^2|\alpha(1-\alpha)]^2}
\end{equation}
and where we used the symmetrization, valid under the respective integral sign,
\begin{equation}
k_{\bar\mu}k_{\bar\nu}k_\mu k_\nu \rightarrow \frac{k^4}{24}\{g_{\bar\mu\bar\nu}g_{\mu\nu}+permutations\}
\label{sym}
\end{equation}
and $\lambda_c=2m^2/(\pi M_{Pl}^2)$.
The integral $I_0$, with the use of the mass counter-term, then leads
us to evaluate the difference,
\begin{equation}
\Delta I=I_0(q)-I_0(0)\cong \frac{\int_0^1 d\alpha \int_0^\infty dx}{2(2\pi)^4}\frac{x^3(x+1)^{-\lambda_c x}}{(x+1)^2(x+1+\bar d)^2}\left(-2\bar d(x+1)-\bar{d}^2\right)
\end{equation}
where we define here $\bar d=|q^2|\alpha(1-\alpha)/m^2$.
It is seen that the dominant part of the integrals comes
from the regime where $x\sim 1/(\rho\lambda_c)$ with $\rho= -\ln\lambda_c$,
so that we may finally write
\begin{equation}
\begin{split}
\Delta I&=I_0(q)-I_0(0)\\
&\cong \frac{\int_0^1 d\alpha \int_0^\infty dx}{2(2\pi)^4}\frac{x^3(x+1)^{-\lambda x}}{(x+1)^2(x+1+\bar d)^2}\left(-2\bar d(x+1)-\bar{d}^2\right)\\
&\cong -\frac{|q|^2I_1}{6(2\pi)^4}-\frac{|q|^4I_2}{60(2\pi)^4}
\end{split}
\label{a22}
\end{equation}
where we have defined
\begin{equation}
\begin{split}
I_1(\lambda_c)&=\int^{\infty}_0dx x^3(1+x)^{-3-\lambda_c x},\nonumber\\
I_2(\lambda_c)&=\int^{\infty}_0dx x^3(1+x)^{-4-\lambda_c x}.
\end{split}
\end{equation}
The result (\ref{a22}) has been used in the text.
\par
For the limit in practice, where we have $\lambda_c\rightarrow 0$, we can get
accurate estimates for the integrals $I_1,I_2$ as follows.
Consider first $I_2$. Write $x^3=(x+1-1)^3= (x+1)^3-3(x+1)^2+3(x+1)-1$
to get
\begin{equation}
\begin{split}
I_2(\lambda_c)&=\int^{\infty}_0dx \left((1+x)^{-1}-3(x+1)^{-2}+3(x+1)^{-3}-(x+1)^{-4}\right)(1+x)^{-\lambda_c x}\nonumber\\
&\cong \int^{\infty}_0dx(x+1)^{-1-\lambda_c x} -\frac{11}{6}.
\end{split}
\end{equation}
Use then the change of variable $r=\lambda_c x$ to get, for $\rho=\ln(1/\lambda_c)$,
\begin{equation}
\begin{split}
 \int^{\infty}_0dx(x+1)^{-1-\lambda_c x}&= \int^{\infty}_0dr\frac{e^{-r\ln(r+\lambda_c)-\rho r}}{r+\lambda_c}\\
&=-\ln\lambda_c+\int^{\infty}_0dr\ln(r+\lambda_c)(\ln(r+\lambda_c)+r/(r+\lambda_c)+\rho)e^{-r\ln(r+\lambda_c)-\rho r}\\
&\cong \rho + \int^{\infty}_0dr\sum_{j=0}^\infty\frac{1}{j!}((\rho +1)(\partial/\partial\alpha)^{j+1}+ (\partial/\partial\alpha)^{j+2})(\partial/\partial\rho)^j r^\alpha e^{-\rho r}|_{\alpha=0}\\
&=\rho +\sum_{j=0}^\infty\frac{1}{j!}((\rho +1)(\partial/\partial\alpha)^{j+1}+ (\partial/\partial\alpha)^{j+2})(\partial/\partial\rho)^j\Gamma(\alpha+1)\rho^{-\alpha-1}|_{\alpha=0}\\
&\cong \rho +\frac{-(\rho+1)\ln\rho +\ln^2\rho}{\rho-\ln\rho}\\
&= \rho -\ln\rho -\frac{\ln\rho}{\rho-\ln\rho}.
\end{split}
\end{equation}
This gives us the approximation
\begin{equation}
I_2(\lambda_c)= \rho -\ln\rho -\frac{\ln\rho}{\rho-\ln\rho} - \frac{11}{6}
\end{equation}
when $\lambda_c\rightarrow 0$, as we noted
in the text.
\par
The integral $I_1$ is a field renormalization constant so, in the usual
renormalization program, we do not need it for most of the applications.
Here, we will discuss it as well for completeness. We get
\begin{equation}
\begin{split}
I_1(\lambda_c)&= \int_0^\infty dx(1+x)^{-\lambda_c x}-3\left(I_2(\lambda_c)+\frac{11}{6}\right)+\frac{5}{2}\nonumber\\
&= \int_0^\infty dx(1+x)^{-\lambda_c x}-3I_2(\lambda_c)-3,
\end{split}
\end{equation}
where, as above, we use
\begin{equation}
\begin{split}
\int_0^\infty dx(1+x)^{-\lambda_c x}&=\frac{\int_0^\infty dr}{\lambda_c}e^{-r\ln(r+\lambda_c)-r\rho}\nonumber\\
&\cong \frac{\int_0^\infty dr}{\lambda_c}\sum_{j=0}^\infty\frac{1}{j!}(\partial/\partial\rho)^j(\partial/\partial\alpha)^jr^\alpha e^{-\rho r}|_{\alpha=0}\nonumber\\
&= \frac{1}{\lambda_c} \sum_{j=0}^\infty\frac{1}{j!}(\partial/\partial\rho)^j(\partial/\partial\alpha)^j\Gamma(1+\alpha)\rho^{-\alpha-1}|_{\alpha=0}\nonumber\\
&\cong  \frac{1}{\lambda_c}\frac{1}{\rho-\ln\rho}.
\end{split}
\end{equation}
Thus, we get
\begin{equation}
I_1(\lambda_c)\cong \frac{1}{\lambda_c}\frac{1}{\rho-\ln\rho} - 3I_2(\lambda_c)-3.
\label{iapprx1}
\end{equation}
\par
Finally, let us show why we can neglect the terms $\bar d$ that were in the denominators of $I_j,~j=1,2$. It is enough to look into the differences
\begin{equation}
\Delta I_j=\frac{\int_0^\infty dx x^3}{(x+1)^j}\left(\frac{1}{(x+1)^2}-\frac{1}{(x+1+\bar d)^2}\right)(x+1)^{-\lambda_c x},~j=1,2
\label{diffints}
\end{equation}
where we note that the integral $I_1$ is absorbed by the standard field renormalization where here for convenience we do this at $|q^2|=0$ when we neglect
$\bar d$ in the denominator of $I_1$ or at the zero of the respective
graviton propagator away from the origin otherwise.
From this perspective, the main integral to examine to illustrate
the level of our approximation becomes
\begin{equation}
\begin{split}
\Delta I_{2}&=\frac{\int_0^\infty dx }{(x+1)^2}\{\frac{(x+1)^{-\lambda_c x}}{(x+1)^2}-\frac{(x+1)^{-\lambda_c x}}{(x+1+\bar d)^2}\}\\
&=\frac{\int_0^\infty dr\, e^{-r\ln(r+\lambda_c)-r\rho}}{(r+\lambda_c)^2}\{\frac{1}{(r+\lambda_c)^2}-\frac{1}{(r+\lambda_c+\sigma)^2}\}\\
&\cong \int_0^\infty dr\int_0^\infty d\alpha_1\alpha_1\int_0^\infty d\alpha_2\alpha_2 e^{-r\ln r-r\rho-\alpha_1(r+\lambda_c)-\alpha_2(r+\lambda_c)}\left(1-e^{-\alpha_2\sigma}\right),
\end{split}
\label{idel2int-1}
\end{equation}
where we have defined $\sigma=\lambda_c\bar d$. The approximation, valid for small values of $\sigma$,
\begin{equation}
\begin{split}
\left(1-e^{-\alpha_2\sigma}\right)&=2e^{-\alpha_2\sigma/2}\sinh(\alpha_2\sigma/2)\\
&\cong \alpha_2\sigma e^{-\alpha_2\sigma/2}
\end{split}
\label{idel2int-2}
\end{equation}
then allows us to get
\begin{equation}
\begin{split}
\Delta I_{2}
&\cong 4\sigma\frac{\partial^2}{\partial\sigma^2}\int_0^\infty dr\, e^{-r\rho}\left(
1-\frac{\lambda_c+\sigma/2}{r+\lambda_c+\sigma/2}\right)\\
&\cong 2+\rho\sigma+2\rho\sigma(1+\frac{1}{4}\rho\sigma)e^{\rho\sigma/2}(C+\ln(\rho\sigma/2)+\sum_{n=1}^{\infty}\frac{(-1)^n(\rho\sigma/2)^n}{n\;n!})\\
\end{split}
\label{idel2int-3}
\end{equation}
which shows that this difference is indeed non-leading log.
The analogous analysis holds for $\Delta I_1$ as well.
\par

\end{document}